\documentclass
[tightenlines,floats,floatfix,aps,notitlepage,nofootinbib,superscriptaddress,11pt]{revtex4-2}
\usepackage[paper=letterpaper,top=32mm, bottom=32mm, left= 23mm, right=23mm]{geometry}
\usepackage[utf8]{inputenc}
\usepackage[table,dvipsnames]{xcolor}
\usepackage{multirow}
\usepackage{graphicx}
\usepackage{amsmath,amssymb,amsfonts,amsthm,latexsym,stmaryrd,physics,mathrsfs}
\usepackage{mathtools}

\allowdisplaybreaks[4]
\usepackage{marginnote}
\usepackage{color}
\usepackage{bm}
\usepackage{longtable}
\usepackage{tikz}
\usepackage{float}
\usepackage{nicefrac}
\usepackage{microtype} 
\usepackage{booktabs}
\usepackage{csquotes}
\usepackage[ruled,linesnumbered]{algorithm2e}

\usepackage{verbatim}
\usepackage{setspace}
\usepackage[T1]{fontenc}
\usepackage[utf8]{inputenc}
\usepackage{stfloats}
\usepackage{diagbox}
\usepackage{dsfont}
\usepackage[colorinlistoftodos,prependcaption,textsize=tiny]{todonotes}
\usepackage{epstopdf}
\usepackage{latexsym}
\usepackage{makecell}
\usepackage{xparse}
\usepackage{accents}
\usepackage[center]{subfigure}
\usetikzlibrary{positioning, shapes.geometric}
\usepackage{xargs}                      
\usepackage[colorinlistoftodos,prependcaption,textsize=tiny]{todonotes}

\newcommand{\hook}{\text{\large{$\lrcorner$}}}
\newcommand{\utilde}[1]{\underaccent{\tilde}{#1}}

\newcommand{\di}{\mathrm{d}}
\usepackage{tensor}
\newcommand{\ou}[3]{\tensor{#1}{^{#2}_{#3}}}
\newcommand{\uo}[3]{\tensor{#1}{_{#2}^{#3}}}

\newcommand{\I}{\mathrm{i}} 

\newcommand{\R}{\mathbb{R}}

\newcommand{\eref}[1]{(\ref{#1})}

\newcommand{\barlambda}{{\mkern0.8mu\mathchar'26\mkern -9.75mu\lambda}}

\DeclareMathAlphabet{\bbgreek}{U}{bbold}{m}{n}

\newcommand{\mtext}[1]{\text{\it #1}}
\usepackage{tikz-cd}

\newcommand\vpm{\mathbin{\vcenter{\hbox{
  \oalign{\hfil$\scriptstyle+$\hfil\cr
          \noalign{\kern-.3ex}
          $\scriptscriptstyle({-})$\cr}}}}}
\DeclareMathAlphabet{\sfit}{OT1}{fos}{sb}{it}
\DeclareMathAlphabet{\mathsf}{OT1}{fos}{sb}{n}

\definecolor{darkgreen}{rgb}{0.01, 0.75, 0.24}
\definecolor{darkblue}{HTML}{2B66D3}

\usepackage[multiple, flushmargin]{footmisc}
\let\originalleft\left
\let\originalright\right
\renewcommand{\left}{\mathopen{}\mathclose\bgroup\originalleft}
\renewcommand{\right}{\aftergroup\egroup\originalright}

\usepackage{scalerel}
\newcommand{\parall}{{\stretchrel*{\parallel}{\perp}}}

\newcommand{\dbarvar}{{\mathrm{d}\mkern-7.5mu\lower.18ex\hbox{$\textasciitilde$}\mkern-1.5mu}}

\DeclareMathAlphabet{\bbvar}{U}{BOONDOX-ds}{m}{n}
\DeclareMathAlphabet{\mathcalvar}{U}{BOONDOX-cal}{m}{n}

\usepackage{ulem}
\renewcommand{\emph}[1]{{\it #1}}
\usepackage{upgreek}

\makeatletter
\def\l@subsection#1#2{}
\def\l@subsubsection#1#2{}
\makeatother
\newcommand{\ac}{A}

{\xspace}
{\xspace}

{\xspace}
{\xspace}

{\xspace}
{\xspace}

\usepackage[colorlinks=true,
            linkcolor=black,
            urlcolor=darkblue,
            citecolor=darkblue]{hyperref}
\begin{document}

\title{Linking Edge Modes and Geometrical Clocks in Linearized Gravity}
\author{Kristina Giesel}
\email{kristina.giesel@fau.de}
\affiliation{Institute for Quantum Gravity, Theoretical Physics III, Department of Physics\\Friedrich-Alexander-Universität Erlangen-Nürnberg, Staudtstra\ss e 7, 91052 Erlangen, Germany\\
}
\author{Viktoria Kabel}
\email{vkabel@ethz.ch}
\affiliation{Institute for Quantum Optics and Quantum Information (IQOQI),\\
Austrian Academy of Sciences, Boltzmanngasse 3, A-1090 Vienna, Austria
}
\affiliation{University of Vienna, Faculty of Physics, Vienna Doctoral School in Physics and\\
 Vienna Center for Quantum Science and Technology (VCQ), Boltzmanngasse 5, A-1090 Vienna, Austria
}
\affiliation{Institute for Theoretical Physics, ETH Zurich, 8093 Switzerland}
\author{Wolfgang Wieland}
\email{wolfgang.wieland@fau.de}
\affiliation{Institute for Quantum Gravity, Theoretical Physics III, Department of Physics\\Friedrich-Alexander-Universität Erlangen-Nürnberg, Staudtstra\ss e 7, 91052 Erlangen, Germany\\
}

\begin{abstract}\vspace{0.5em}
\noindent Reference frames are crucial for describing local observers in general relativity. In quantum gravity, we have different proposals for how to understand them. There are models in which the reference frames remain classical. In other approaches, they are fundamentally quantum. Recently, two options appeared to investigate these possibilities at the level of the classical and quantum algebra of observables. One option is based on the covariant phase space approach, using gravitational edge modes. In the canonical approach, there is another option, relational clocks, built from matter or geometry itself. In this work, we extend existing results and show how to relate edge modes and geometrical clocks in linearized gravity. We proceed in three steps.  First, we introduce an extension of the ADM (Arnowitt--Deser--Misner) phase space to account for covariant gauge fixing conditions and the explicit time dependence they add to Hamilton's equations. Second, we show how these gauge fixing conditions recover a specific choice of geometrical clocks in terms of Ashtekar--Barbero connection variables. Third, we study the effect of the Barbero--Immirzi parameter on the generators of asymptotic symmetries. We introduce the corresponding charges and explain how this parameter, which disappears from metric gravity, modifies the generators for angle-dependent asymptotic symmetries. This modification affects the super-translation charges, while the global charges remain unchanged.
\end{abstract}
\maketitle

\vspace{-2em}
\tableofcontents
\makeatletter
\let\toc@pre\relax
\let\toc@post\relax
\makeatother 

\newpage

\section{Introduction}
\noindent Formulating a physical model often requires a choice of reference frame. This becomes particularly important in systems that involve gravity as well as for systems that are fundamentally quantum. In general relativity (GR), reference frames are  linked to the diffeomorphism invariance of the theory. For quantum systems, the question arises whether reference systems are fundamentally classical or quantum. In quantum gravity, these two aspects of the problem are no longer separable.\smallskip

The study of dynamical reference frames in the context of general relativity is motivated by the aim of formulating the dynamics of gauge invariant quantities and closely linked with the problem of time \cite{Isham_1992,Kuchar_1999,Anderson_2010}. Since GR describes a fully constrained system, the canonical Hamiltonian vanishes in the physical sector. All observables are required to have vanishing Poisson brackets with it and thus evolve trivially under the canonical Hamiltonian. This problem can be resolved by selecting a suitable subset among the fundamental kinematical degrees of freedom as a reference frame. In the relational formalism \cite{Rovelli_1991_observable,Rovelli_1994,Rovelli_2002_partialobservables, Dittrich_2004, Dittrich_2005, Vytheeswaran_1994, Thiemann_2004, Hoehn_2019_trinity}, we can then understand the evolution of the remaining degrees of freedom relative to this dynamical and physical reference frame. The overall objective is often to recast the resulting evolution equations into a Hamiltonian form generated by a \emph{physical Hamiltonian}. Unlike the canonical Hamiltonian of GR, the physical Hamiltonian is not just a linear combination of constraints and does not vanish in the physical sector of the theory. Furthermore, it generates a non-trivial relational evolution of the physical observables relative to the chosen reference frame. A common choice for such a reference frame is matter fields. Such matter reference frames have been built from dust fields, dating back to the seminal work in \cite{Kuchar_1990,Brown_1994}, or from suitable sets of scalar fields as in \cite{Kuchar_1991,Rovelli_1993,Kuchar_1995} with many developments in the context of loop quantum gravity \cite{Giesel_2007,Domagala_2010,Lewandowski_2011,Husain_2011,Giesel_2012,Giesel_2016} in particular. There is also the option to use embedding fields as part of an extended phase space \cite{Isham_1984a,Isham_1984b}. Another possibility is to consider reference frames which are constructed from the gravitational degrees of freedom alone, so-called geometrical clocks. This is a particularly convenient choice for vacuum spacetimes or perturbations around them. Proposals for such geometrical clocks date back to the early days of general relativity, see for example \cite{kretschmann_1918, Komar_1958}. They were further discussed in the early works on the Hamiltonian formulation of GR \cite{Arnowitt_1962,Kuchar_1970}, and have, for instance, been applied more recently in \cite{Dittrich_2007, Giesel_2018, Fahn_2022}. Note that, in these more recent works, the geometric clocks of \cite{Arnowitt_1962,Kuchar_1970} were extended to include the additional Gauss constraint that is present when formulating GR in terms of Ashtekar--Barbero variables.\footnote{The extensions chosen in \cite{Dittrich_2007} and \cite{Fahn_2022,Fahn_2024} differ slightly and a comparison is discussed in the appendix of \cite{Fahn_2022}.}\smallskip

Reference frames also emerge naturally in the context of bounded subregions in GR as \emph{edge modes} for the diffeomorphism group \cite{Balachandran_1994b, Carlip_1996, Donnelly_2016, Donnelly_Giddings_2016, Gomes_2016, Geiller_2017a, Speranza_2017, Geiller_2017b, Wieland_2017a, Wieland_2017b, Takayanagi_2019, Freidel_2019, Francois_2020, Freidel_2020_em1, Freidel_2020_em2, Freidel_2020_em3, Donnelly_2020, Wieland_2020, Freidel_2021_extended, Ciambelli_2021b, Wieland_2021a, Carrozza_2022, Goeller_2022, Ciambelli_2022b, Freidel_2023}. While gauge degrees of freedom are typically considered irrelevant redundancies in our description of physical systems, they do carry physical significance when considering subsystems or bounded subregions. By restricting ourselves to a finite region, we can consider large gauge transformations that change the field configuration at the boundary. Such transformations are then no longer a redundancy in our description of the system. The boundary modes that carry a representation of these large boundary gauge symmetries encode otherwise missing information, crucial for gluing adjacent regions back together, see \cite{Rovelli_2013_whygauge, Rovelli_2020_gauge, Gomes_2019}. In modern treatments, following \cite{Donnelly_2016}, edge modes are typically studied within the covariant phase space formalism \cite{Lee_1990, Iyer_1994, Wald_1999, Harlow_2019}. There, the emergence of additional boundary degrees of freedom can be understood in terms of the symplectic structure of the classical phase space. Under large gauge transformations, the symplectic two-form changes by a boundary term, which determines the phase space of the gauge group-valued reference fields and their conjugate momenta at the boundary of the spacetime region. In the case of GR, the relevant symmetries are diffeomorphisms. The corresponding edge modes take the form of coordinate or embedding fields, depending on the formalism. This observation already suggests a close connection between edge modes and reference frames, which is made more explicit in \cite{Carrozza_2021,Carrozza_2022,Goeller_2022,Kabel_2023}, where it is shown that edge modes can serve as dynamical reference frames in gauge theory and gravity. At the quantum level, edge modes can be related to quantum reference frames (QRFs) as studied in the quantum foundations and quantum information literature (see, e.g.,~\cite{Giacomini_2017_covariance, Loveridge_2017, Vanrietvelde_2018a, Hoehn_2019_trinity, delaHamette_2020,  delaHamette_2021_perspectiveneutral, Castro_Ruiz_2021, Carette_2023, Kabel_2024} for modern approaches).\smallskip

In this work, we investigate the link between two specific choices of reference frames: \emph{geometrical clocks} \cite{kretschmann_1918, Komar_1958, Arnowitt_1962,Kuchar_1970, Dittrich_2007, Giesel_2018, Fahn_2022} and \emph{edge modes} \cite{Balachandran_1994b, Carlip_1996, Donnelly_2016, Donnelly_Giddings_2016, Gomes_2016, Geiller_2017a, Speranza_2017, Geiller_2017b, Wieland_2017a, Wieland_2017b, Takayanagi_2019, Freidel_2019, Francois_2020, Freidel_2020_em1, Freidel_2020_em2, Freidel_2020_em3, Donnelly_2020, Wieland_2020, Freidel_2021_extended, Ciambelli_2021b, Wieland_2021a, Carrozza_2022, Goeller_2022, Ciambelli_2022b, Freidel_2023}. While both of these choices of reference frame have been applied to the same physical regime, namely linearized gravity \cite{Fahn_2022, Kabel_2023}, and do not draw upon additional matter degrees of
freedom to construct the reference frames, they differ substantially in the mathematical framework. The study of geometrical clocks is largely conducted within the canonical formulation of general relativity and quantum gravity, whereas edge modes are usually described in the covariant phase space formalism. This raises the question whether we can establish an explicit relation between the two. The aim of this paper is to answer this question by linking the canonical and covariant formulations employed in this research area.\smallskip

More concretely, we focus on the relation between geometrical clocks and edge modes in the regime of linearized gravity, as studied in \cite{Fahn_2022, Kabel_2023}. We develop the connection between these different types of reference frames in several steps. First, starting from the canonical description, we propose a new extension of the ADM phase space that enables the formulation of covariant gauge fixing conditions. Second, beginning with a covariant phase space description, we incorporate the Barbero--Immirzi parameter, used in recent definitions of geometrical clocks \cite{Dittrich_2007, Giesel_2018, Fahn_2022} in terms of Ashtekar--Barbero variables. Finally, we demonstrate that covariant gauge-fixing conditions imposed at the level of the covariant phase space reproduce the geometrical clocks in the canonical formalism for a specific partial gauge fixing. Thus, the geometrical clocks of \cite{Fahn_2022} can be seen as a specific choice of the coordinate reference fields of \cite{Kabel_2023}. In other words, the latter constitute a gauge-unfixed version of the former. The gauge fixing condition that we will use to establish this link is the harmonic gauge fixing condition. In \cite{Kuchar_1991}, it was shown how the harmonic gauge can be implemented in  the canonical theory using matter reference frames constructed from scalar fields. Here, we are primarily interested in a different problem. Our goal is to link  the harmonic gauge condition to geometric clocks rather than matter reference frames. Therefore, we must follow an alternative strategy here. Our strategy is more closely related to \cite{Froeb_2017}, where the harmonic gauge condition is used to express the coordinates in terms of metric perturbations, in order to compute gauge-invariant correlation functions for a scalar field.\smallskip

Our research deepens our understanding of the connection between edge modes and dynamical reference frames previously explored in \cite{Carrozza_2021,Carrozza_2022,Goeller_2022,Kabel_2023}. While  in \cite{Carrozza_2021,Carrozza_2022} the edge modes arise from the complement of the subregion in spacetime, we consider here the subregion as primary. Moreover, we focus on the context of linearized general relativity and determine an explicit mapping between two specific choices of reference frames in this regime. Since, from our perspective, these reference frames are chosen with the goal of formulating models of quantum gravity in mind, the algebra of the observables associated with these reference frames plays an important role. At the non-linear level, this algebra generally has a complicated structure that confronts us with a challenging problem of representation theory in the context of canonical quantization. The algebra simplifies drastically in the linearized theory. In this regime, some of the technical challenges can be avoided and our specific choice of reference fields poses no obstacles to quantization.\smallskip

Since our work lies at the intersection of several different areas of research---the relational formalism in general relativity and quantum gravity, the study of boundary modes and their associated charges, and the field of quantum reference frames---we also discuss how our results relate to earlier research in the various communities involved. First, we provide a comparison of our extension of the ADM phase space with previous results reported in \cite{Pons_1996,Pons_1999,Garcia_2000,Pons_2003,Kouletsis_2008}. We discuss several differences between the elementary phase space variables, highlighting that our construction not only facilitates the implementation of generic gauge fixing conditions, but also allows to generate the time evolution of otherwise frozen Lagrange multipliers through a corresponding extended Hamiltonian. Second, we compute the effect of the Barbero--Immirzi parameter on the boundary charges in the covariant phase space formalism. We show that while $\gamma$ has no effect on the global Poincaré charges---the gravitational energy and momentum when taking the subregion to infinity---it can alter higher multipoles, obtained by integrating the charges non-uniformly over the asymptotic two-sphere. Third, we discuss the potential effects for the resulting quantum reference transformations. While the generators of global translations remain unaffected, the Barbero--Immirzi parameter can affect the generators of angle-dependent translations and thereby also the corresponding quantum reference frame transformations.\smallskip

The paper is structured as follows: After the introduction in Section \ref{sec:ADMphasespace}, we start from the canonical formalism and construct an extended ADM phase space, which facilitates the formulation of covariant gauge fixing conditions. The covariant phase space formulation is the topic of  Section \ref{sec:covariantPS}, in which we add the Barbero--Immirzi parameter in the derivation of the edge modes, placing particular focus on the linearized regime. In Section \ref{sec:coordConditions}, we then demonstrate how the covariant harmonic gauge condition reproduces the geometrical clocks of the canonical formalism under an appropriate partial gauge fixing for the additional variables in the extended phase space. Section \ref{sec:relationOtherWork} embeds our work in the broader literature, by relating it to previous extensions of the ADM phase space (Subsection \ref{subsec:relADMphaseSpace}), demonstrating the effects of the Barbero--Immirzi parameter on the boundary charges in the covariant phase space formalism (Subsection \ref{sec:boundaryCurrents}), and outlining the connection to quantum reference frame transformation (Subsection \ref{sec:BIparameter}). We close in Section \ref{sec:Conclusions} with a concluding discussion and an outlook on future work. 
\vfill
\pagebreak

\newgeometry{top=32mm, bottom=15mm, left= 23mm, right=23mm}

\subsection*{Notation}

\setstretch{0.5}
\begin{align*}
    \mu,\nu,\dots \hspace{4em} &\text{four-dimensional tensor indices}\\
    a,b,\dots \hspace{4em} &\text{three-dimensional tensor indices}\\
    \alpha,\beta,\dots \hspace{4em} &\text{four-dimensional internal indices}\\
    i,j,\dots \hspace{4em} &\text{three-dimensional internal indices}\\
    \mathcal{M} \hspace{4em} &\text{spacetime manifold}\\
    \mathcal{\mathbb{M}} \hspace{4em} &\text{flat target space of the coordinate fields}\\
    \Sigma \hspace{4em} &\text{Cauchy hypersurface}\\
    g_{\mu\nu} \hspace{4em} &\text{four-metric}\\
    h_{ab} \hspace{4em} &\text{three-metric}\\
    d^4v_g \hspace{4em} &\text{metrical volume element on }\mathcal{M}\\
    d^3v_h \hspace{4em} &\text{metrical volume element on }\Sigma\\
    \ast \hspace{4em} &\text{internal Hodge operator}(\ast X_{\alpha\beta} = \tfrac{1}{2}\epsilon_{\alpha\beta}^{\;\;\;\; \alpha' \beta'}X_{\alpha' \beta'})\\
    \epsilon_{\mu\nu\rho\sigma} \hspace{4em} &\text{Levi Civita tensor on $\mathcal{M}$ } (\epsilon_{0123} = \sqrt{|\det g|}) \\
    \epsilon_{abc} \hspace{4em} &\text{Levi Civita tensor on $\Sigma$ } (\epsilon_{123} = \sqrt{\det h})\\
    \tilde{\epsilon}_{abc} \hspace{4em} &\text{Levi Civita tensor density on $\Sigma$ } (\utilde{\epsilon}_{123} = 1)\\
    \tilde{\pi}^{ab} \hspace{4em} &\text{ADM canonical momentum}\\
    K_{ab} \hspace{4em} &\text{extrinsic curvature}\\
    X^\mu \hspace{4em} &\text{coordinate fields}\\
    \tilde{P}_\mu \hspace{4em} &\text{conjugate momenta to }X^\mu\\
    t_\mu \hspace{4em} &\text{time-flow co-vector field}\\
    \tilde{\Pi}^\mu \hspace{4em} &\text{conjugate momentum to }t_\mu\\
    n^\mu \hspace{4em} &\text{normal vector to }\Sigma\\
    \tilde{p}_\mu \hspace{4em} &\text{conjugate momentum to }n^\mu\\
    N \hspace{4em} &\text{lapse function}\\
    N^a \hspace{4em} &\text{shift vector}\\
    \tilde{\mathcal{H}} \hspace{4em} &\text{Hamiltonian constraint}\\
    \tilde{\mathcal{H}}_a \hspace{4em} &\text{spatial diffeomorphism (vector) constraint}\\
    e_\alpha^{~\mu} \hspace{4em} &\text{tetrad}\\
    e_i^{~a} \hspace{4em} &\text{triad}\\
    \tilde{E}_i^{~a} \hspace{4em} &\text{densitized triad } (\tilde{E}_i^{~a} = \sqrt{\det(h)}\uo{e}{i}{a})\\
    A^\alpha_{\;\;\beta} \hspace{4em} &\text{spin connection in $\mathcal{M}$}\\
    \omega^i_{\;\;j} \hspace{4em} &\text{pull-back of spin connection onto } \Sigma\\
    A^i_{\;\; a} \hspace{4em} &\text{Ashtekar-Barbero connection}\\
     \kappa \hspace{4em} &\text{gravitational coupling constant }(\kappa = 16 \pi G)\\
    \Omega_\Sigma \hspace{4em} &\text{pre-symplectic form}\\
    \theta \hspace{4em} &\text{symplectic potential}\\
    \bbvar{d} \hspace{4em} &\text{exterior derivative on field space }\\
    \bbvar{D} \hspace{4em} &\text{covariant derivative (dressed variation) on field space}\\
    ^{(0)}\Phi^I \hspace{4em} &\text{background configurations are labeled with $^{(0)}$, here for $\Phi^I$}\\
    \delta \Phi^I \hspace{4em} &\text{first order perturbations are denoted by $\delta$, here for $\Phi^I$}
\end{align*}

\vfill
\pagebreak 
\restoregeometry
\section{Reference Frames in the Context of the  (Extended) ADM Phase Space}
\label{sec:ADMphasespace}
\noindent 
There exist two ways to deal with gauge degrees of freedom in a gauge theory. Either we can choose certain gauge fixing conditions, apply the gauge-fixing and consider gauge-fixed theory. The second option is to link the gauge fixings with reference frames and formulate the corresponding relational dynamics in terms of gauge-invariant quantities, or, in other words, observables. In the latter case, we can translate a given gauge fixing into a choice of reference systems with respect to which observables and their relational dynamics can then be obtained. When we consider the later quantization of a theory, we have the option of either carrying out the above steps at the classical level and then quantising only the physical degrees of freedom, known as reduced phase space quantization, or quantising all kinematical degrees of freedom and extracting the physical sector of the theory and its dynamics at the quantum level, denoted as Dirac quantization.

To be able to relate reference frames chosen in the canonical theory to those chosen in the covariant formalism and thus to map a particular choice of reference fields from the level of the covariant phase space into the canonical formalism, we need to impose a covariant gauge fixing on the canonical ADM (Arnowitt--Deser--Misner) phase space and vice versa. In this section, we will introduce the necessary tools to perform this step. For this purpose, we first introduce an extension of the ADM phase space in Subsection \ref{sec:ExtADM}. This extension is strongly motivated by the goal of implementing covariant gauge fixing conditions in the canonical theory. Second, we apply this framework to our problem. Since there are other extensions of the ADM phase space available, in Subsection \ref{subsec:relADMphaseSpace} below, we compare our present proposal with other  available extensions in the literature.

\subsection{Extension of the ADM phase space}\label{sec:ExtADM}
\noindent To formulate the extension of the ADM  phase space \cite{Arnowitt_1962} we start from the Einstein--Hilbert action with boundary terms 
\begin{align}
S_{\mathrm{EH}}[g_{\mu\nu}]&=\frac{1}{\kappa}\left[\int_{\mathcal{M}}d^4v_g\,R+2\oint_{\partial\mathcal{M}}d^3v_h\,\epsilon K\right],\label{EHactn-def}
\end{align}
where $\kappa:=16\pi G$, with $G$ being Newton's constant, $g_{\mu\nu}$ denotes the spacetime metric on $\mathcal{M}$, the Ricci scalar is $R$, $K$ is the trace of the extrinsic curvature and with $\epsilon=\pm 1$ where the normal to the boundary $\partial\mathcal{M}$ denoted by $n_B$ satisfies $n_B^\mu n^B_\mu=\epsilon$. 
In what follows, we need the action as a functional on the reduced ADM phase space, where the lapse function $N$ and the shift vector $N^a$ are treated as Lagrange multipliers. We obtain this action upon introducing a $3+1$ decomposition of the spacetime manifold. Assuming ${\cal M}\simeq \Sigma\times \mathbb{R}$, we obtain the bulk action
\begin{align}
S[h_{ab},\tilde{\pi}^{ab};N,N^a]=\frac{1}{\kappa}\int\!\di t\int_{{\Sigma}}\di^3x\Big[\tilde{\pi}^{ab}\dot{h}_{ab}-N\Big(\utilde{G}_{abcd}\tilde{\pi}^{ab}\tilde{\pi}^{cd} - \sqrt{\det(h)}{}^{(3)}R\Big)-2\tilde{\pi}^{ab}D_aN_b\Big].\label{eq:EHADM}
\end{align}
Notice that $\mu,\nu,\dots$ are abstract tensor indices in $\mathcal{M}$, raised (lowered) with the metric tensor $g^{\mu\nu}$  ($g_{\mu\nu}$), whereas $a,b,c,\dots$ are abstract tensor indices on the spatial submanifold $\Sigma\subset\mathcal{M}$. Spatial indices are raised (lowered) using the spatial three-metric $h^{ab}$ ($h_{ab}$).
To obtain the action in \eqref{eq:EHADM}, we followed the standard ADM procedure, in which we use the Gauss--Codazzi equation
\begin{equation}
R = {}^{(3)}R+(K_{ab}K^{ab}-K^2)-2\nabla_\mu(\beta^\mu-n^\mu K),
\end{equation}
that relates the Ricci scalar in four dimensions to the one on the spatial manifold $\Sigma$, 
where $K_{ab}$ is the extrinsic curvature and $\beta^\mu$ denotes the acceleration of the normal vector $n^\mu$ to $\Sigma$, each given by
\begin{align}
K_{ab}&=\di Y^\mu_{~a}\di Y^\mu_{~b}\nabla_\mu n_\nu=\frac{1}{2N}\left(\dot{h}_{ab}-2 D_{(a}N_{b)}\right),\label{eq:Kab-def}\\
\label{eq:BetaAcc}
 \beta^\mu&=n^\nu\nabla_\nu n^\mu.
\end{align}
In  \eref{eq:Kab-def}, we introduced  embedding maps $\di Y^\mu_{~a}$ which are $T{\cal M}$-valued one-forms on $\Sigma$. We construct them as follows. If $\{Y^\varrho\}$ is a local coordinate chart of $\mathcal{M}$ that overlaps with a local coordinate chart ${ x}^r$ on $\Sigma$, the components of the embedding maps $\ou{\di Y}{\mu}{a}$ are defined by
\begin{equation}
\ou{\di Y}{\mu}{a}\left[\frac{\partial}{\partial x^r }\right]^a=\frac{\partial(Y^\varrho\circ x^{-1})}{\partial  x^r }\left[\frac{\partial}{\partial Y^\varrho }\right]^\mu\in T\mathcal{M}\big|_\Sigma.\label{eq:Y-solder-def}
\end{equation}
In addition, we introduced the Wheeler--De Witt super-metric $\utilde{G}_{abcd}$, i.e.\ an inverse tensor  density on $\Sigma$, and its inverse, which is a tensor density of weight one,
\begin{align}
\tilde{G}^{abcd}&= \sqrt{\det h}\left(h^{a(c}h^{d)b}-h^{ab}h^{cd}\right),\\
\utilde{G}_{abcd}&=\frac{1}{\sqrt{\det h}}\left(h_{a(c}h_{d)b}-\tfrac{1}{2}h_{ab}h_{cd}\right).
\end{align}
The Legendre transformation replaces the velocities of the spatial metric $\dot{h}_{ab}$ by their canonical ADM momenta
\begin{equation}
\tilde{\pi}^{ab}=\tilde{G}^{abcd}K_{cd}=\sqrt{\det(h)}(K^{ab}-h^{ab} K).\label{eq:AMD-p-def}
\end{equation}
The resulting kinematical phase space is vastly bigger than the physical phase space that describes the two radiative modes of gravitational waves. One way to access these modes is to use appropriate gauge fixing conditions---so their solutions exist. On the covariant phase space, we may use covariant gauge conditions, such as, for example, harmonic gauge, that is $\Box_g X^\mu=0$. This equation imposes a condition on four scalar fields $X^\mu$, intended to serve as reference frames, as well as on some of the metric components. There is a dependence on  the metric, simply because the condition involves the d'Alembertian $\Box_g=g^{\mu\nu}\nabla_\mu\nabla_\nu$. On the other hand, when working on the ADM reduced phase space, we can find geometrical clocks, which impose conditions on the components of the spatial metric  $h_{ab}$  and their conjugate momenta $\tilde{\pi}^{ab}$. When translating covariant gauge fixing conditions such as $\Box_gX^\mu=0$ into the reduced ADM phase space two problems arise. First, not all metric components are dynamical on the ADM phase space. The lapse function $N$ and the shift vector $N^a$ are Lagrange multipliers rather than variables on phase space. Second, the covariant gauge fixing condition is usually applied to reference frames that play the role of embedding fields $X^\mu:\Sigma\rightarrow\R^4$ on the canonical side, see for instance \cite{Isham_1984b,Isham_1984a,Kouletsis_2008}. Lapse and shift, on the other hand, are related to the normal and tangential projection of the velocities $\dot{X}^\mu$ of these embedding fields. At the ADM Hamiltonian level, it is difficult to make sense of such time derivatives of covariant reference fields. Now, the first problem can be circumvented by working with an extended phase space introduced in \cite{Pons_2003,Pons_2009,Pons_2010}. In this formalism, lapse and shift are promoted into phase space variables with corresponding canonical momenta extending the usual ADM variables $(h_{ab}, \tilde{\pi}_{ab})$. This solution was developed in \cite{Lee_1990,Pons_1996,Pons_1999,Garcia_2000,Pons_2009,Pons_2010}. However, once we also take the second problem into account, such an extension is no longer convenient from the perspective of this work, since the four-dimensional history of lapse and shift is required to recover the embedding fields from them. For this reason, we choose a different extension of the ADM phase that we introduce in the following.

As mentioned above, our extension of the ADM phase space is guided by the idea to implement covariant gauge fixing conditions at the canonical level. To elaborate on this in more detail, we consider that a  physical reference frame is defined in a relational way. In the Lagrangian theory, a simple choice is to use harmonic coordinates $X^{\varrho}:\mathcal{M}\longrightarrow\mathbb{M}$ that satisfy the wave equation
\begin{equation}
\square_gX^{\varrho}=g^{\mu\nu}\nabla_\mu\nabla_\nu X^{\varrho} =0,\label{eq:gaugecond1}
\end{equation}
with embedding fields $X^{\varrho}$ taking values in an unspecified target space $\mathbb{M}$, e.g.\ $\mathbb{M}=\R^4\ni(X^0,X^1,$ $X^2,$ $X^3)$. If we then want to solve the gauge conditions \eref{eq:gaugecond1}, we need to specify initial data
\begin{equation}
X^{\varrho}\big|_{\Sigma_o},\quad n^\mu\nabla_\mu X^{\varrho}\big|_{\Sigma_o} =n^{\varrho}
\end{equation}
on some initial surface $\Sigma_o$, where $n^\mu\in T\mathcal{M}$ is the future pointing time-like normal vector to $\Sigma_o$. To make the construction covariant, two further choices must be made. First, we need to choose a coordinate invariant condition to select the hypersurface $\Sigma_o$. We could say, for instance, that it defines an extremal surface, i.e.\ a surface at which the trace of the extrinsic curvature vanishes, i.e.\ $K=\ou{K}{a}{a}=0$. Second, we need to provide initial data on the then selected hypersurface $\Sigma_o$ using, once again, some coordinate invariant prescription. A simple possibility is, for example, to require that $n^\mu\nabla_\mu X^\varrho|_{\Sigma_o}=\delta^\varrho_0$, $X^0\big|_{\Sigma_o}=0$, and $\Delta_hX^i=h^{ab}D_aD_bX^i=0$ such that we only have to fix the asymptotic falloff conditions\footnote{In here, $r $ is the radial coordinate $r=\sqrt{\delta_{ij}X^i X^j}\rightarrow\infty$ and $\hat{r}^i$ is the radial unit vector with respect to the background metric $\delta_{ab}=\delta_{ij}\partial_a X^i\partial_b X^j$ on $\Sigma_o$.} to determine $X^i$ for given $h_{ab}$ on $\Sigma_o$. Using a perturbative expansion around a flat background, $h_{ab}=\delta_{ab}+\delta h_{ab}+\mathcal{O}(\delta^2)$, we could then solve $X^i=r \hat{r}^i+\delta Q^i(\vartheta,\varphi)+\mathcal{O}(\delta^2)$ order by order in $\delta h_{ab}$. The next to leading order $\delta Q^i(\vartheta,\varphi)$ of the reference frame will have a non-trivial functional dependence on the linearized ADM three-metric on $\Sigma_o$, schematically $\delta Q^i(\vartheta,\varphi)\equiv\delta Q^i [\delta h_{ab}](\vartheta,\varphi)$.
From the perspective of the canonical ADM formalism, the residual corner data $\delta Q^i(\vartheta,\varphi)$, which is now a functional on the ADM kinematical phase space, is a good candidate for the edge modes that appear in the covariant phase space approach. We will justify this idea below in Section \ref{sec:BIparameter}. By integrating the gauge condition \eref{eq:gaugecond1}, we  then obtain a reference frame $X^\varrho$ in a neighborhood of $\Sigma_o$. This reference frame depends as a functional on the metric and as an ordinary function on the points of the spacetime manifold. This functional will be covariant, i.e.\ it satisfies for any diffeomorphism $\varphi\in\mathrm{Diff}(\mathcal{M})$ that
\begin{equation}
X^\varrho[\varphi^\ast g_{\mu\nu}]=X^\varrho[g_{\mu\nu}]\circ\varphi.
\end{equation}
There are two main drawbacks with this construction. First, we cannot expect the so-defined coordinates to cover all of spacetime, let alone to separate points on a sufficiently large class of solutions to Einstein's equations. Second, the construction becomes highly non-local and non-linear when we try to make sense of it on the reduced ADM phase space. As mentioned above, in this phase space, there is no longer a spacetime metric tensor available. We only have the cotangent bundle of spatial geometries parametrized by ADM data $(h_{ab},\tilde{\pi}^{ab})$ available. To translate the covariant reference frames $X^\varrho[g_{\mu\nu}]$ into a functional on the reduced ADM phase space, we would need to take the exponential of the Hamiltonian vector field and evolve our initial data $(h_{ab},\tilde{\pi}^{ab})$ from the $K=0$ hypersurface $\Sigma_o$ back into a generic configuration of ADM data on a generic hypersurface $\Sigma$. This seems unpractical: solving Hamilton's equations is equally challenging as solving Einstein's field equations at the full non-perturbative level. A strategy to implement the harmonic gauge fixing condition in the canonical theory is presented in \cite{Kuchar_1991}.
In this construction, a different extended phase space was used than the one we will introduce here. In  \cite{Kuchar_1991}, which builds upon the seminal work \cite{Isham_1984b,Isham_1984a}, the phase space consists of the ADM variables $(h_{ab},\tilde{\pi}^{ab})$, 
four additional scalar fields, which  serve as reference fields, and their respective momenta. In addition, there are Lagrange multipliers for lapse and shift and four more scalar fields that serve as Lagrange multipliers that impose the harmonic gauge condition for the four reference fields. The harmonic gauge fixing condition is then implemented at the level of the action. This action is a functional of eight additional scalar fields besides the metric. The system has first-class and second-class constraints. Upon imposing the second-class constraints, one ends up with a system of first-class constraints that has four additional massless scalar fields. In this model, these four additional massless scalar fields, which are minimally coupled to gravity, serve as matter reference frames.
Such a matter reference frame   plays a double role. It induces a foliation of spacetime and provides embedding fields, mapping kinematical observables into Dirac observables. {This procedure amounts to a non-perturbative dressing, c.f.\ \cite{Giddings_2019}.} One of our aims is to link edge modes with the geometrical clocks in \cite{Fahn_2022}. Such geometrical clocks are built directly from the ADM data. Therefore, the strategy developed in \cite{Kuchar_1991} to implement the harmonic gauge condition through a matter reference frame cannot be applied in our case. Instead, we use the harmonic gauge condition directly. This allows us to relate the coordinate fields to the metric degrees of freedom directly, similar to the construction in \cite{Froeb_2017} but now formalized on a new extension of the ADM phase space.

To proceed, we will work on an enlarged phase space. Furthermore, we will not implement the harmonic gauge condition at the level of the action. Instead, consider first the 3+1 split of the covariant gauge fixing condition \eref{eq:gaugecond1}. We obtain
\begin{align}
\square_gX^\varrho=-(n^\mu\nabla_\mu)^2X^\varrho+\beta^aD_aX^\varrho-Kn^\mu\nabla_\mu X^\varrho+\Delta_h X^\varrho,\label{eq:gaugecond2}
\end{align}
where $\Delta_h $ is the Laplacian of $h_{ab}$ and $\beta^a$ is the acceleration \eref{eq:BetaAcc}, which can be written as
\begin{equation}
\beta_a={\di Y}_{\mu a}n^\nu\nabla_\nu n^\mu=N^{-1}D_aN,\label{eq:acclrtn}
\end{equation}
with $N$ denoting the lapse function. One of the main problems is now manifest. While the differential operator $\Delta_h$ and the trace of the extrinsic curvature $K$ can be written in terms of the elementary ADM variables, this is not possible for $n^\mu\nabla_\mu$ and $\beta^a=h^{ab}D_b\ln N$. At this stage, they are treated as mere $c$-numbers that Poisson commute with all phase space variables. Neither the lapse function nor the normal derivative $n^\mu\nabla_\mu$ exist as operators or functionals defined on the reduced ADM phase space. Thus, the reduced ADM phase space, i.e.\ the cotangent bundle of the superspace of spatial three-metrics, is too small to accommodate the covariant gauge fixing \eref{eq:gaugecond2} as a local constraint thereon. 

To resolve this issue, we introduce additional configuration variables and corresponding conjugate momenta. First, we add the reference fields $X^\mu$ to the extended phase space, and introduce a corresponding densitized momentum variable $\tilde{P}_\mu$, which is a $T^\ast\mathbb{M}$-valued three-form on $\Sigma$, i.e.\
\begin{equation}
X^\mu:\Sigma\rightarrow\mathbb{M},\quad \tilde{P}_\mu\in \Omega^3(\Sigma:T^\ast \mathbb{M}).
\end{equation}
Next, we have to take care of normal derivatives of $X^\varrho$ on the enlarged phase space. To express  $n^\mu\nabla_\mu X^\varrho\equiv n^\varrho(X)$ on the extended phase space, we introduce another canonical pair
\begin{equation}
n^\mu\in \Omega^0(\mathbb{M}:T\mathbb{M}),\quad \tilde{p}_\mu\in \Omega^4(\mathbb{M}:T^\ast\mathbb{M}).
\end{equation}
Finally, we need to also express the lapse function, which appears in the covariant gauge condition \eref{eq:gaugecond2} through the acceleration vector $\beta^a=N^{-1}D^aN$, as a functional on the extended phase space. The lapse function is the normal component with respect to $\Sigma$ of 
the time-flow vector field denoted by $t^\mu$ and given by
\begin{equation}
t^\mu=Nn^\mu+N^\mu,\quad N^\mu:=\ou{\di Y}{\mu}{a}N^a.
\end{equation}
For reasons that will become clear later, it is useful to rearrange the multipliers $N$ and $N^a$ into a target space covector field $t_\mu$ with canonical conjugate momenta $\widetilde{\Pi}^\mu$,
\begin{equation}
t_\mu\in \Omega^0(\mathbb{M}:T^\ast\mathbb{M}),\quad \widetilde{\Pi}^\mu\in \Omega^4(\mathbb{M}:T\mathbb{M}).
\end{equation}
We equip the extended ADM phase space with a natural symplectic potential for these variables
\begin{align}
\Theta=\frac{1}{\kappa}\int_{\Sigma}\tilde{\pi}^{ab}\bbvar{d}h_{ab}+\int_\Sigma\tilde{P}_\mu\bbvar{d}X^\mu+\int_{\mathbb{M}}\left(\tilde{\Pi}^\mu\bbvar{d}t_\mu+\tilde{p}_\mu\bbvar{d}n^\mu\right).
\end{align}
 The  non-vanishing Poisson brackets on the extended ADM phase space read
\begin{align}
\left\{h_{ab}(q),\tilde{\pi}^{cd}(q')\right\}&=\kappa\delta^{c}_{(a}\delta^{d}_{b)}\tilde{\delta}^{(3)}_\Sigma(q,q')\label{eq:PoissADM}\\
\left\{X^{\mu}(q),\tilde{P}_\nu(q')\right\}&=\delta^{\mu}_\nu\tilde{\delta}^{(3)}_\Sigma(q,q')\label{eq:Poiss1},\\
\left\{t_{\mu}(x),\tilde{\Pi}^\nu(x')\right\}&=\delta_{\mu}^\nu\tilde{\delta}^{(4)}_{\mathbb{M}}(x,x'),\label{eq:Poiss2}\\
\left\{n^{\mu}(x),\tilde{p}_\nu(x')\right\}&=\delta_{\nu}^\mu\tilde{\delta}^{(4)}_{\mathbb{M}}(x,x').\label{eq:Poiss3}
\end{align}
Note that the first two terms define Poisson brackets that are local on $\Sigma$, which can be understood here as an abstract 3-dimensional hypersurface, whereas the last two terms are local on the target space $\mathbb{M}$, which is 4-dimensional.

If we now take the flat  metric $\eta_{\mu\nu}$ on the target space $\mathbb{M}$, we can split the three-metric $h_{ab}$ into a background plus a perturbation
\begin{equation}
h_{ab}=\delta_{ab}+2f_{ab},\qquad \delta_{ab}=\eta_{\mu\nu}\partial_aX^{\mu} \partial_bX^{\nu}.
\end{equation}
In an asymptotically flat spacetime, we can choose coordinates such that the perturbation  $f_{ab}$ falls off as $\rho^{-1}$ for $\rho=\sqrt{\eta_{\mu\nu} X^\mu X^\nu}\rightarrow\infty$. It is possible to relax these conditions allowing for a weaker falloff, see \cite{Fiorucci:2024ndw} for recent results on this issue.   
If we include the boundary terms at the asymptotic two-sphere, the canonical Hamiltonian $\boldsymbol{H}$ on the extended ADM phase space has the following form:
\begin{align}
\label{eq:Hdef}
\boldsymbol{H}&=\int_\Sigma\left[-n^\mu(X)t_\mu(X)\left(\tilde{\mathcal{H}}+n^{\nu}(X)\tilde{P}_{\nu}\right)+D^aX^\mu t_\mu(X)\left(\tilde{\mathcal{H}}_a+D_aX^{\nu}\tilde{P}_{\nu}\right)\right]\\
&\quad+\frac{1}{8\pi G}\oint_{\partial\Sigma}d^2v_a\left(-n^\mu(X)t_\mu(X)\partial_b(f^{ab}-\delta^{ab}\ou{f}{c}{c})+D_bX^\mu t_\mu(X){\pi}^{ab}\right),\nonumber
\end{align}
where $\partial_a$ is the covariant derivative for the background metric $\delta_{ab}$ and $d^2v_a$ is the directed area element at the boundary, i.e.\ a co-vector valued two-form with components $(d^2v_a)_{bc}=\varepsilon_{abc}$. Furthermore, we defined 
\begin{align}
\kappa\,\tilde{\mathcal{H}}&=\utilde{G}_{abcd}\tilde{\pi}^{ab}\tilde{\pi}^{cd}-\sqrt{\det(h)}\,{}^{(3)}R \label{eq:Ham-cons}\\
\kappa\, \tilde{\mathcal{H}}_a &= -2D_b\ou{\tilde{\pi}}{b}{a}\label{eq:Vec-cons}
\end{align}
The Hamiltonian is a functional $\boldsymbol{H}[h_{ab},\tilde{\pi}^{ab},X^\mu,\tilde{P}_\mu,n^\mu,t_\mu]$ on the extended ADM phase space. The corresponding action on the extended ADM phase space has 12 primary constraints 
\begin{equation}
\tilde{P}_\mu\approx 0,\quad\widetilde{\Pi}^\mu\approx 0,\quad\tilde{p}_\mu\approx 0.\label{eq:primry}
\end{equation}
The stability of the primary constraints, that is 
$\{ \tilde{P}_\mu(q),\boldsymbol{H}\}\stackrel{!}{=}0$ and likewise for the remaining primary constraints yields the secondary Hamiltonian constraint and the spatial diffeomorphism or vector constraint 
\begin{equation}
\tilde{\mathcal{H}}=0,\qquad \tilde{\mathcal{H}}_a=0.\label{eq:secndry}
\end{equation}
Since the secondary constraints commute with all primary constraints and satisfy the ADM hypersurface deformation algebra, there are no further constraints. In total, the system has 16 first class constraints on the extended ADM phase space. Considering this, we have 18 kinematical degrees of freedom\footnote{These are represented by 36 phase space dimensions.} in the extended ADM phase space and two physical degrees of freedom. This is, of course, the same number of physical degrees of freedom that we have in the reduced ADM phase space for vacuum GR.

Given the extended canonical Hamiltonian $\boldsymbol{H}$ we can write down Hamilton's equations for the set of canonical variables $(h_{ab},\tilde{\pi}^{ab};X^\mu,\tilde{P}_\mu;t_\mu,\tilde{\Pi}^\mu;n^\mu,\tilde{p}_\mu)$, i.e.\
\begin{align}
\frac{\di}{\di t} h_{ab}&=\left\{h_{ab},\boldsymbol{H}\right\}\equiv\mathfrak{X}_{\boldsymbol{H}}[h_{ab}],\\
\frac{\di}{\di t} \tilde{\pi}^{ab}&=\left\{\tilde{\pi}^{ab},\boldsymbol{H}\right\}\equiv\mathfrak{X}_{\boldsymbol{H}}[\tilde{\pi}^{ab}]
\end{align}
and likewise for the other canonical variables where $\mathfrak{X}_{\boldsymbol{H}}=\{.,\boldsymbol{H}\}$ denotes the Hamiltonian vector field of $\boldsymbol{H}$. The full set of Hamilton's equations together with the 12 primary and four secondary constraints is equivalent to Einstein's equations in the Lagrangian formulation.

\begin{figure}[h]
\centering        
\includegraphics[scale=0.24]{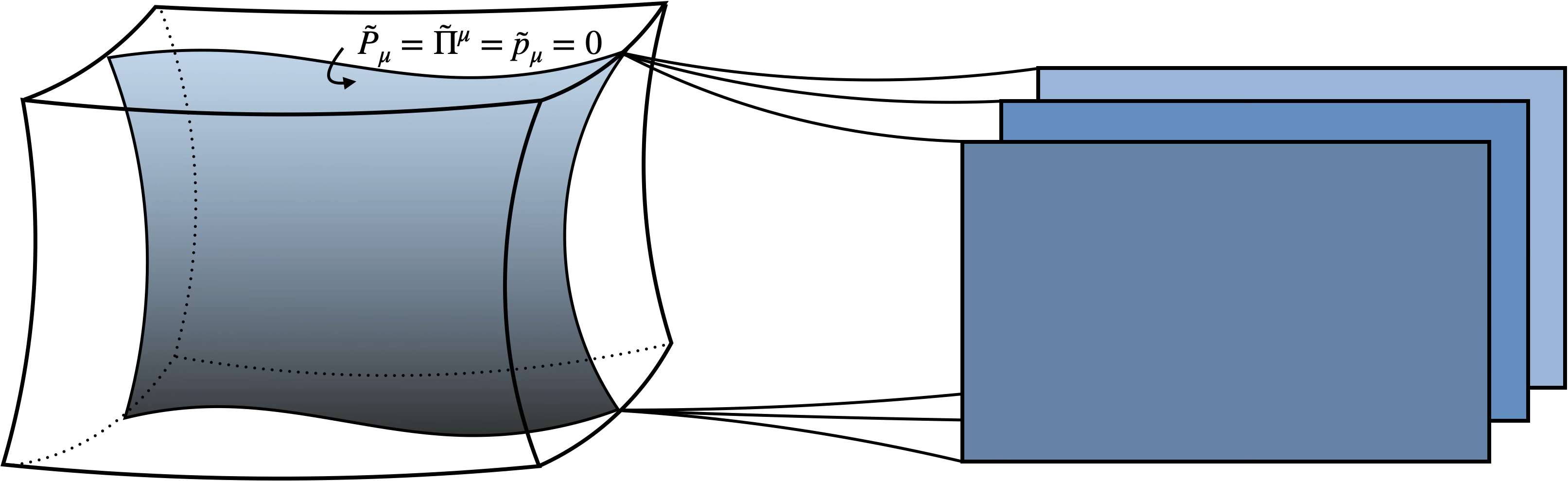}
    \caption{The extended phase space constructed in this work is represented by the three-dimensional object on the left-hand side. Each point is a configuration $(h_{ab},\tilde{\pi}^{ab};X^\mu,\tilde{P}_\mu; t_\mu,\tilde{\Pi}^\mu; n^\mu,\tilde{p}_\mu )$. By performing a symplectic reduction with respect to the constraints $\tilde{P}_\mu, \tilde{\Pi}^\mu,$ and $\tilde{p}_\mu$, we obtain the reduced ADM phase space, where each point is a configuration of the gravitational degrees of freedom  $(h_{ab},\tilde{\pi}^{ab})$ only. It is depicted as the blue embedded surface on the left. The different colors represent its additional dimensions, fanned out on the right.}
    \label{fig:RedToADM}
\end{figure}

A symplectic reduction with respect to the 12 primary constraints listed in \eqref{eq:primry} recovers the reduced ADM phase space with only $(h_{ab},\tilde{\pi}^{ab})$ as elementary phase space variables. On this phase space, we have the  corresponding Hamilton's equations, as well as the four secondary constraints, namely the Hamiltonian and spatial diffeomorphism constraint \eref{eq:secndry}. See Figure \ref{fig:RedToADM} for an illustration.

The lapse function as well as the shift vector can be expressed on the extended ADM phase space as
\begin{equation}
N=-n^\mu(X)t_\nu(X),\qquad N^a = h^{ab}\partial_bX^\mu t_\mu(X)=h^{ab}N_b.\label{eq:NN-def}
\end{equation}
Hence, after the symplectic reduction with respect to all primary constraints, the auxiliary variables $n^\mu, t_\mu$ and $X^\mu$ are no longer dynamical variables such that $N$ and $N_a$, that is lapse and shift, become Lagrange multipliers in the reduced ADM phase space.

Given the extension of the ADM phase space described above, we have achieved our first intermediate result. We have a phase space at hand, which allows us to write down the covariant harmonic gauge condition as a not too complicated yet local functional on the extended phase space. As illustrated in Figure \ref{fig:GF} below, the next step ahead is to understand the relation between edge modes and geometric clocks. We will achieve this by choosing appropriate gauge fixing conditions for the 16 first-class constraints $\tilde{\cal H},\tilde{\cal H}_a, \tilde{P}_\mu,\tilde{p}^\mu, \widetilde{\Pi}_\mu$. For this purpose, we select a second-class constraint partner for each of the first-class constraints, or, more precisely, for combinations thereof. Of course, such a choice is not unique, but for a given fixed choice, the order in which the individual constraints are implemented is weakly equivalent. However, the choice of the second-class partners for the first-class constraints defines a specific choice of gauge fixings, which can, in turn, be related to a specific choice of reference frame. Different gauge conditions lead to different descriptions of the same physical system with respect to different reference frames. Not all such descriptions are equally useful. Depending on how tractable it is to construct the resulting algebra of Dirac observables, some reference frames may be be better than others. This becomes particularly important when considering the canonical quantization of the physical sector of the model.

Instead of considering the Hamiltonian and diffeomorphism constraint directly, we consider the following four combinations of equivalent first class constraints
\begin{equation}
\label{eq:DefHmu}
{\boldsymbol{H}}^\mu :=\frac{\delta {\boldsymbol{H}}}{\delta t_\mu(X)} = g^{\mu\nu}\tilde{P}_\nu -n^\mu(X)\tilde{\mathcal{H}}+D^aX^\mu\tilde{\mathcal{H}}_a
,
\end{equation}
where $t_\mu(X)(q):=(t_\mu\circ X)(q)$ and we defined the inverse metric on $\Sigma$ with respect to the reference frame $X^\mu:\Sigma\rightarrow\mathbb{M}$ in target space via
\begin{equation}
\label{eq:mtrc-def}
g^{\mu\nu}(X)\equiv g^{\mu\nu}[h_{ab},X^\mu,n^\mu]=-n^\mu(X)n^\nu(X)+h^{ab}D_aX^\mu D_bX^\nu.
\end{equation}
For a suitable choice of gauge fixing conditions for $n^\mu$ and $X^\mu$ it is possible to identify ${\boldsymbol{H}}^0$ and ${\boldsymbol{H}}^a$ with the Hamiltonian and spatial diffeomorphism constraint in the reduced ADM phase space if we additionally require $\tilde{P}_\mu=0$ to be satisfied. 
In addition, we have the remaining twelve first class constraints that set $\tilde{P}_\mu,\tilde{p}^\mu, \widetilde{\Pi}_\mu$ to zero.
As shown in Figure \ref{fig:GF}, we perform the gauge fixing in the following way:
\begin{itemize}
\item[(i)] First we implement the gauge fixing $\Phi^\mu=0$ as the second class partner for the constraints $\boldsymbol{H}^\mu$.
\item[(ii)] As a second step, we choose a gauge fixing for both $n^\mu$ and $t_\mu$ as a second-class partner for $\tilde{\Pi}^\mu$ and $\tilde{p}^\mu$ respectively. 
\item [(iii)] Finally, we impose a gauge fixing condition for $X^\mu$ as a second-class partner for $\tilde{P}_\mu$.
\end{itemize}

\begin{figure}[h]
\centering
\subfigure[Extended Phase Space]{
        \centering
       \includegraphics[scale=0.25]{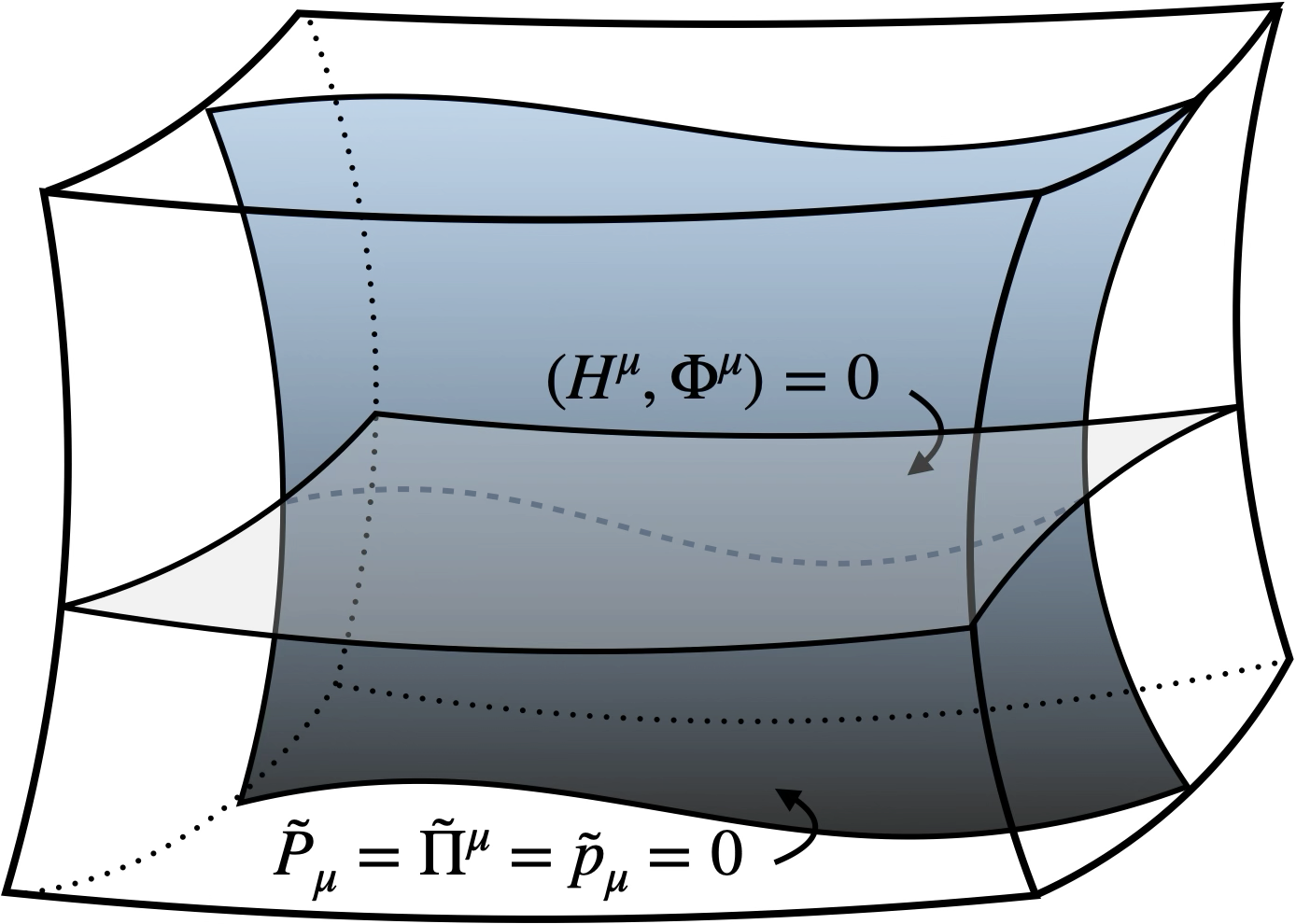}
       \label{fig:extended_PS2}
   } \qquad
    \subfigure[Reduced ADM Phase Space]{
        \centering
        \includegraphics[scale=0.25]{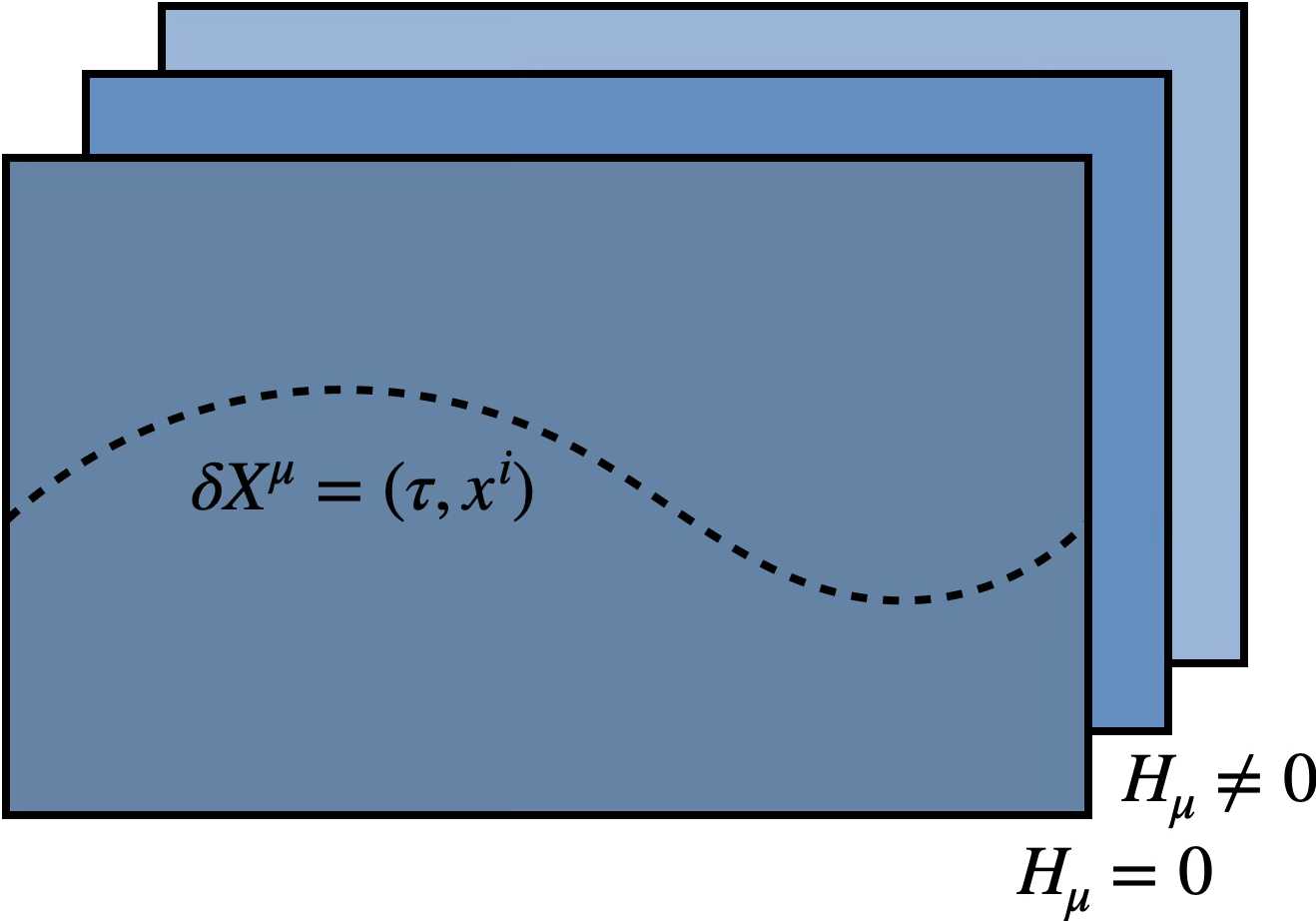}
        \label{fig:ADM2}
    }
    \caption{(a) On the level of the \emph{extended phase space}, we first implement the harmonic gauge fixing condition $\Phi^\mu =\square_g X^\mu = 0$, which relates the coordinate fields to the gravitational degrees of freedom, as the second class partner for the constraints $\boldsymbol{H}^\mu = 0$ (gray horizontal surface). To return to the ADM phase space, we additionally have to impose the constraints $\tilde{P}_\mu =  \tilde{\Pi}^\mu = \tilde{p}_\mu =0$ and impose appropriate gauge fixing conditions on their conjugate partners $n^\mu, t_\mu,$ and $X^\mu$ (blue vertical surface). The intersection of these two gauge-fixing surfaces reproduces the gauge-fixing obtained by fixing the values of a particular set of geometrical clocks $\delta X^\mu = (\tau, x^i)$ and setting the constraint $\boldsymbol{H}_\mu$ to zero on the level of the \emph{ADM phase space} (b), as depicted by the dotted line on the $\boldsymbol{H}_\mu =0$ hypersurface.}
    \label{fig:GF}
\end{figure}

Let us elaborate a little further on why we perform this type of gauge fixing, why we do it in this specific order, and to what kind of reference systems the whole procedure corresponds. First of all, we note that the gauge fixings for $X^\mu, t_\mu$ and $n^\mu$ all include one of the configuration variables linearly in addition to some functions on either ${\cal M}$ or $\mathbb{M}$. In contrast,  the harmonic gauge fixing condition $\Phi^\mu=\Box_g X^\mu=0$ is a functional of several variables, namely $X^\mu, t_\mu, n^\mu, h_{ab}, \tilde{\pi}^{ab}$, which reads on the extended phase space
\begin{eqnarray}
\label{eq:gaugecond3}
\Phi^\mu &=&-(\partial_{\nu} n^\mu)(X)n^{\nu}(X)+(\partial_{\rho} \ln N)(X)D^aX^{\rho} D_aX^\mu -\frac{h_{ab}\tilde{\pi}^{ab}}{2\sqrt{\det(h)}}n^\mu(X)+\Delta_hX^\mu=0.
\end{eqnarray}
The three second-class pairs in (ii) and (iii) each remove four degrees of freedom (eight phase space dimensions) from the extended ADM phase space. In this way, we can successively remove the canonical pairs $(X^\mu,\tilde{P}^\mu), (t_\mu,\widetilde{\Pi}^\mu)$ and $(n^\mu,\tilde{p}_\mu)$ from the extended phase space. Going back to (i), we will be left to impose the remaining gauge fixing condition $\Phi^\mu=0$. At this stage, it will be  a condition on the gravitational degrees of freedom alone, removing 8 phase space dimensions from the set of canonical variables $(h_{ab},\tilde{\pi}^{ab})$. Notice however that the condition $\Phi^\mu=0$ depends on the gauge fixing conditions for the other first-class constraints $\tilde{\Pi}^\mu=0$, $\tilde{p}^\mu=0$ and $\tilde{P}^\mu=0$. As we will see in the further part of this research, this observation is crucial. It provides exactly what we need to relate edge modes and geometrical clocks. This also explains why we reverse the order and perform the harmonic gauge fixing condition first (as indicated in the list above). Since we now impose $\Phi^\mu=0$ first, we obtain correlations between the ADM data  $(h_{ab},\tilde{\pi}^{ab})$ and the auxiliary configurations of $X^\mu, t_\mu$ and $n^\mu$. Thus, the constraint $\Phi^\mu =0$ provides a more general gauge-fixing on the extended phase space and thereby also a more general reference frame than the geometrical clock variables that have been constructed in \cite{Fahn_2022}. However, as we will demonstrate below, for a particular simple choice of gauge fixing conditions for 
$X^\mu,t_\mu$ and $n^\mu$ the two constructions agree on the reduced ADM phase space. In this sense, we will understand the extended phase space introduced here, together with the harmonic gauge fixing condition $\Phi^\mu=0$, as the gauge-unfixed version of the geometric clocks constructed in \cite{Fahn_2022}. The equivalence will be demonstrated explicitly below, when comparing the two frameworks at the level of linearized gravity. 

In the following, let us first briefly discuss how the geometrical clocks can in principle also be chosen as reference frames at the full non-linear level. This is possible in principle, but there are are also practical challenges that are drastically simplified when  considering the linearized theory.

At the non-linear level, the Poisson bracket of the second class pair $(\Phi^\mu,{\boldsymbol{H}}^\nu)$ is given by
\begin{equation}
\{\Phi^\mu(q),{\boldsymbol{H}}^{\nu}(q')\} = \Box_g \{X^\mu(q),{\boldsymbol{H}}^\nu(q')\} + \{\Box_g, {\boldsymbol{H}}^\nu(q')\}X^\mu(q).
\end{equation}
To compute the contribution of the first Poisson bracket, we consider the smeared version of the constraints ${\boldsymbol{H}}^{\nu}$ and the generator $\boldsymbol{H}_\xi$
\begin{equation}
\boldsymbol{H}_\xi:=\int\limits_{\Sigma}\xi_\mu(X) \frac{\delta\boldsymbol{H}}{\delta t_\mu(X)}
\label{eq:Hxi-def}
\end{equation}
where $\xi_\mu(q)=(\xi_\mu\circ X)(q)$ is a co-vector field on $\Sigma$.
Then we have 
\begin{equation}
\label{eq:Hxi-X}
\{X^\mu, \boldsymbol{H}_\xi\} = (g^{\mu\nu}\xi_\nu)(X)
\end{equation}
and from this we immediately obtain
\begin{equation}
\{X^\mu(q), \boldsymbol{H}^{\nu}(q')\} = g^{\nu\rho}(X)\delta^{\mu}_\rho\tilde{\delta}^{(3)}_\Sigma(q,q').
\end{equation}
Furthermore, for a generic metric $g_{\mu\nu}$ also the second contribution involving $\{\Box_g, {\boldsymbol{H}}^\nu(q')\}$ will be non-vanishing so that in the non-linear theory we have
\begin{equation}
\{\Phi^\mu(q),{\boldsymbol{H}}^{\nu}(q')\}\not=0. 
\end{equation}
However, for $\Phi^\mu$ to be a suitable second-class partner for ${\boldsymbol{H}}^{\nu}$, the Dirac matrix $M^{\mu\nu}(q,q'):=\{\Phi^\mu(q),{\boldsymbol{H}}^{\nu}(q')\}$ must satisfy $\det(M^{\mu\nu})(q,q')\not=0$ to be invertible. In the case of a gauge fixing, this is necessary to construct the corresponding Dirac bracket. In the relational formalism, this condition must be fulfilled. Otherwise, it is not possible to construct the corresponding Dirac observables with respect to a reference frame defined by $\Phi^\mu$. Moreover, the algebra of these observables can be shown to be weakly equivalent to \cite{Thiemann_2004}
\begin{equation}
 \{ {\cal O}_f\, ,\, {\cal O}_g\} \simeq {\cal O}_{\{f,g\}^*_{(\Phi^\mu,{\boldsymbol{H}}^\nu)}}   
\end{equation}
where we denote the Dirac observable of a function $f$ by ${\cal O}_f$ and $\{\cdot,\cdot\}^*_{(\Phi^\mu,H^\nu)}$ denotes the Dirac bracket associated with the symplectic reduction of the pair of second class constraints $(\Phi^\mu,{\boldsymbol{H}}^\nu)$. 

At the non-linear level, two problems arise. The first problem is that inverting the matrix $(M^{\mu\nu})(q,q')$ normally also requires inverting the d'Alembert operator $\Box_g$. This in turn means that the corresponding Green's function  must exist for generic $g_{\mu\nu}$. Yet, even if we were able to invert $(M^{\mu\nu})(q,q')$, we will be left with a further second problem. At the non-linear level, for most of the elementary phase space variables on the extended phase space, the Dirac bracket $\{\cdot,\cdot\}^*_{(\Phi^\mu, {\boldsymbol{H}}^\nu)}$ does not coincide with the Poisson bracket. In most cases at least either the configuration variables or the momenta of the canonical variables will not commute with $\Phi^\mu$ or ${\boldsymbol{H}}^\mu$ or both. For this reason, we end up with an algebra for elementary Dirac observables that do not assume the usual form in terms of standard canonical commutation relations in general. At the classical level, this may not be a problem, but if we aim to find a representation of this algebra at the quantum level, this means that we have to deal with a difficult and non-standard representation problem. Whether suitable representations exist in this case is an open problem.

It is not too surprising that these difficulties occur. In the canonical approach, one finds that the search for non-linear geometric clocks, with respect to which the physical sector of the theory can be quantized, is a daunting  task that remains largely unresolved. To circumvent the problems just discussed, we consider edge modes and geometric clocks in the framework of the linearized theory, as in \cite{Fahn_2022}, where the problems mentioned above do not exist. The fact that these problems are drastically simplified in the linearized theory can already be expected upon noting that the linearized constraints ${\boldsymbol{H}}^\mu$ are Abelian. The details of the linearized theory will be presented in \ref{sec:coordConditions}. However, by discussing the structure of the linearized constraints and gauge fixing conditions, we can already understand here at this stage why it leads to a simplification. 

Consider thus the linearized harmonic gauge fixing condition, i.e.\
\begin{equation}
    \delta\Phi^\mu = {}^{(0)}\Box\delta X^\mu + (\delta \Box_g){}^{(0)}X^\mu,
\end{equation}
where we denote background quantities with a label ${}^{(0)}$ and perturbations with $\delta$. From \eqref{eq:gaugecond3}, we realize that $(\delta \Box_g){}^{(0)}$ depends on the perturbations $\delta n^\mu,\delta t_\mu, \delta h_{ab},\delta\tilde{\pi}^{ab}$ as well as the corresponding background quantities. If we choose a background metric ${}^{(0)}g_{\mu\nu}$, such as for instance the Minkowski metric for which the Green's function of the d'Alembert operator exists, we can solve $\delta \Phi^\mu$ for $\delta X^\mu$ and obtain an equivalent gauge fixing condition that is linear in $\delta X^\mu$. Similarly, from \eqref{eq:DefHmu} we obtain at the linearized level
\begin{equation}
\delta {\boldsymbol{H}}^\mu = {}^{(0)}g^{\mu\nu}\delta\tilde{P}_\nu - {}^{(0)}n^\mu(X)\delta\tilde{\cal H}+{}^{(0)}D_a{}^{(0)}X^\mu\delta\tilde{\cal H}_a,
\end{equation}
where we used that it is possible to choose a background in which the constraints ${}^{(0)}\tilde{P}_\nu, {}^{(0)}\tilde{\cal H}$ and ${}^{(0)}\tilde{\cal H}_a$ are satisfied and thus trivially vanish. Moreover, $\delta {\boldsymbol{H}}^\mu$ can be easily solved for $\delta\tilde{P}_\mu$ yielding an equivalent set of linearized constraints that satisfy
\begin{equation}
 \{{}^{(0)}\Box^{-1}\delta \Phi^\mu(q), {}^{(0)}g_{\nu\rho}\delta H^\rho(q')\} =\delta^{\mu}_\nu\delta_\Sigma^{(3)}(q,q') + \mathcal{O}(\delta)  
\end{equation}
and have the property that the ${}^{(0)}g_{\nu\rho}\delta H^\rho(q')$ are weakly Abelian up to the order in perturbation theory that we are interested in. However, the four components of ${}^{(0)}\Box^{-1}\delta \Phi^\mu(q)$ will not mutually Poisson commute. In  \cite{Fahn_2022}, a set of mutually commuting geometric clocks was obtained by introducing the method of the so-called dual observable map, which leads to a weakly equivalent but extended version of the geometric clocks. Whether this strategy can be generalized to the covariant theory is an interesting question that we will investigate in future work.

Notice also that, at the linearized level, picking a gauge-fixing surface $\delta\Phi^\mu=0$ amounts to imposing a functional dependence between the configurations of  $\delta h_{ab},\delta\tilde{\pi}^{ab}, \delta X^\mu, \delta t_\mu$. For given  $\delta X^\mu,\delta t_\mu$ and $\delta n^\mu$, we can interpret the resulting residual constraints on $(h_{ab},\delta\tilde{\pi}^{ab})$ as the gauge-unfixed version of the gauge-fixing condition associated with the geometrical clocks used in \cite{Fahn_2022}.

One further step is required to show  that for a certain choice of gauge-fixing conditions for $\delta X^\mu,\delta t_\mu$ and $\delta n^\mu$ we can reproduce the exact functional form of the geometrical clocks reported in \cite{Fahn_2022} in this way. What is still missing is to translate our results into first-order spin-connection variables to match the representation of the geometrical clocks in terms of Ashtekar--Barbero variables, see \cite{Fahn_2022}. For this purpose, we consider reference frames for the first-order tetrad variables on the covariant phase space, see Section \ref{sec:covariantPS} below. The results of Section \ref{sec:covariantPS} form the basis for  linking edge modes and geometric clocks in Section \ref{sec:coordConditions}.

\section{Reference Frames in the Covariant Phase Space}
\label{sec:covariantPS}

\noindent In the previous section, we saw how to extend the ADM canonical phase space to impose covariant gauge fixing conditions on the reference fields for the diffeomorphism group. In this section, we turn to the covariant phase space description and show how such reference fields arise at the boundary of spacetime subregions as edge modes for the diffeomorphism group. Since we will relate the covariant reference fields to geometrical clocks constructed from Ashtekar--Barbero variables, we start from the Hilbert--Palatini action with Barbero--Immirzi parameter $\gamma$. In Subsection \ref{sec:symplecticFull}, we derive the corresponding bulk and boundary pre-symplectic two-form for full general relativity. In Subsection \ref{sec:symplecticlinearized}, we restrict ourselves to linearized perturbations around a flat spacetime background. In either case, we find that the resulting boundary terms describe additional degrees of freedom, which provide reference frames for diffeomorphisms and internal Lorentz rotations.

\subsection{Pre-Symplectic Form}\label{sec:symplecticFull}

\noindent The pre-symplectic form is obtained from the variation of the action. We start from the Hilbert--Palatini action with Immirzi parameter $\gamma$,
\begin{equation}
	S[e,A]=\frac{1}{16\pi G}\int_{\mathcal{M}}\left(\ast(e_\alpha\wedge e_\beta)-\frac{1}{\gamma}e_\alpha\wedge e_\beta\right)\wedge F^{\alpha\beta}[A].\label{eq:actndef}
\end{equation}
Our conventions are: $\ou{F}{\alpha}{\beta}=\mathrm{d}\ou{A}{\alpha}{\beta} + \ou{A}{\alpha}{\gamma}\wedge \ou{A}{\gamma}{\beta}$ is the curvature of the $SO(1,3)$ spin connection $\ou{A}{\alpha}{\beta}$ and $\nabla_a[\cdot]=\partial_a[\cdot]+[A,\cdot]$ is the corresponding exterior gauge covariant derivative with curvature two-form $\ou{F}{\alpha}{\beta}=\ou{[\nabla_a,\nabla_a]}{\alpha}{\beta}$. In addition, $e_\alpha$ is the tetrad and $\ast$ denotes the \emph{internal} Hodge operator: $\ast(e_\alpha\wedge e_\beta)=\frac{1}{2}\uo{\epsilon}{\alpha\beta}{\alpha'\beta'}e_{\alpha'}\wedge e_{\beta'}$. Given the action, the pre-symplectic potential $\theta$ is inferred from the variation by
\begin{equation}
	\delta S[\Phi^I] = E_I\delta\Phi^I + \oint_{\partial\mathcal{M}}\theta[\Phi^I,\delta \Phi^I],
\end{equation}
where $\Phi^I$ denotes the dynamical fields. Setting $E_I \approx 0$ imposes the equations of motion. This procedure for inferring $\theta$ from the variation of the action does not fix the symplectic potential uniquely---instead, there are ambiguities arising from the freedom to add an arbitrary field space exact form to $\theta$ and from the freedom to add additional boundary terms to the Lagrangian \cite{Iyer_1994}. See \cite{Freidel_2023,Wieland_2017a,Riello_2020,DePaoli_2018,Freidel_2021_extended,Oliveri_2019}, for further background on the physical relevance of these ambiguities.  Given the action \eqref{eq:actndef}, a possible choice for $\theta$ is given by
\begin{equation}
	\theta(\delta) = \frac{1}{16\pi G}(\ast-\gamma^{-1})(e_\alpha\wedge e_\beta)\wedge\delta A^{\alpha\beta} = \frac{1}{16\pi G}\Pi_{\alpha\beta} \wedge \bbvar{d}A^{\alpha\beta}(\delta),
\end{equation}
where, in the second part, we have introduced the conjugate momentum $\Pi_{\alpha\beta}=(\ast-\gamma^{-1})(e_\alpha\wedge e_\beta)$ as well as an exterior derivative $\bbvar{d}$ on field space, which is related to variations of the dynamical fields (tangent vectors $\delta\Phi^I$ in field space) by $\bbvar{d}\Phi^I(\delta) = \delta \Phi^I$. The Poisson brackets are inferred from the pre-symplectic two-form, which is obtained by taking the exterior field-space derivative of the pre-symplectic potential and integrating the resulting expression over a Cauchy hypersurface $\Sigma$:
\begin{equation}
	\Omega_\Sigma=\frac{1}{16\pi G}\int_\Sigma\bbvar{d}\Pi_{\alpha\beta}\wedge\bbvar{d}A^{\alpha\beta}.\label{eq:Omdef}
\end{equation}

Now, the action \eref{eq:actndef} is invariant under both four-dimensional diffeomorphisms and internal $SO(1,3)$ frame rotations. This leads to a degeneracy in the pre-symplectic two-form $\Omega_\Sigma$. To deal with this redundancy, we introduce covariant reference frames for both groups: $X^\mu:\mathcal{M}\rightarrow\mathbb{M}$ for the diffeomorphisms and  $\ou{\Lambda}{\alpha}{\mu}:\mathbb{M}\rightarrow SO(1,3)$ for the $SO(1,3)$ gauge transformations.\footnote{Note that the internal gauge transformations are themselves expressed relative to a given choice of coordinate fields $X^\mu$, i.e. they take elements of $\mathbb{M}$ rather than $\mathcal{M}$.} These can be used to parametrize the fundamental  variables relative to the reference fields, as
\begin{align}
e^\alpha&=\ou{\Lambda}{\alpha}{\mu}(X)\left(\di X^\mu+\ou{f}{\mu}{\nu}(X)\di X^\nu\right)\equiv\ou{\Lambda}{\alpha}{\mu}(X)\left(\di X^\mu+{f}^{\mu}(X)\right),\label{eq:tetrad}\\
\ou{A}{\alpha}{\beta}&=\ou{\Lambda}{\alpha}{\mu}(X)\di\uo{\Lambda}{\beta}{\mu}(X)+\ou{\Lambda}{\alpha}{\mu}(X)\ou{\Delta}{\mu}{\nu}(X)\uo{\Lambda}{\beta}{\nu}(X).\label{eq:connection}
\end{align}
In fact, as noted in \cite{Kabel_2023}, at the linearized level, such a choice for the reference frames is always possible, because any flat line element determines a set of inertial reference fields $\{X^\mu\}$ and vice versa. In this sense, the reference fields $(\ou{\Lambda}{\alpha}{\mu},X^\mu)$ can be seen as encoding the same information as the flat background tetrad $e^\alpha$. Taking into account variations of the reference frames is the same as making the background metric dynamical. For the time being, we will not impose gauge conditions, such as those considered in Section \ref{sec:ADMphasespace}, on the coordinates $X^\mu$ and Lorentz reference frames $\ou{\Lambda}{\alpha}{\mu}$. We will do so in Section \ref{sec:coordConditions}, where we connect our results to the canonical formulation of \cite{Fahn_2022} in terms of Ashtekar-Barbero variables.

To express the symplectic two-form in terms of the parameterization \eqref{eq:connection} and \eqref{eq:tetrad}, we consider the variation of the tetrad and connection as we change the embedding fields $X^\mu:\mathcal{M}\rightarrow\mathbb{M}$. To simplify our notation, we introduce
\begin{equation}
\bbvar{X}_p=\bbvar{d}X^\mu(p)\left[\frac{\partial}{\partial X^\mu}\right]_p\equiv\bbvar{d}X^\mu\partial_\mu\Big|_p,
\end{equation}
as a tangent vector-valued one-form on field-space to encode variations of the coordinate fields. Notice that $\bbvar{X}_p$ defines for any $p\in\mathcal{M}$ a linear map from the infinite-dimensional tangent space $T_X\mathcal{X}$, where the base manifold $\mathcal{X}$ denotes here the infinite-dimensional space of embedding fields $X^\mu:\mathcal{M}\rightarrow\mathbb{M}$, into the tangent bundle $T_p\mathcal{M}$. Conversely, we can lift any vector field $V\in \mathfrak{X}(\mathcal{M})$ into a vector field $\delta_V\in \mathfrak{X}(\mathcal{X})$ by setting $\delta_V[X^\mu(p)]=V^\mu_p$, where $V^\mu_p$ are the components of the vector field $V\in\mathfrak{X}(\mathcal{M})$ with respect to the tangent basis $\{\partial_\mu\}$ of the coordinate fields $\{X^\mu\}$. The contraction of the vector field $\delta_V\in T\mathcal{X}$ with the one-form $\bbvar{X}$ is then simply given by $\bbvar{X}(\delta_V)=V$. In summary, the map $\mathcal{X}\ni X^\mu\rightarrow \bbvar{X}_p\in T^\ast_X\mathcal{X}\otimes T_pM$ defines a $T_p\mathcal{M}$-valued one-form, i.e.\ a section of $\Omega^1(\mathcal{X}:T_pM)$. This one-form is not an exact form on field space. Instead, it satisfies the Maurer--Cartan equation $\bbvar{d}\bbvar{X} = \frac{1}{2}[\bbvar{X},\bbvar{X}]$, where $[\cdot,\cdot]$ is the Lie bracket between vector fields on spacetime such that e.g.\ $(\bbvar{d}\bbvar{X})(\delta_U,\delta_V)=[U,V]$. Note that there is a similar equation in the BRST approach to the quantization of gauge systems. Indeed, viewed as a one-form, which is naturally anti-commuting, $\bbvar{X}$ behaves like the anti-commuting Fadeev--Popov ghosts for the diffeomorphism group, see e.g.\ \cite{Gomes_2016,Ciambelli_2021b}.

The main practical purpose of introducing the reference fields $\ou{\Lambda}{\alpha}{\mu}$ and $X^\mu$ is to distinguish between \emph{physical} variations of the fundamental fields and variations that can be absorbed by mere \emph{gauge transformations}. To this end, we introduce a covariant derivative on field-space, i.e.\ a dressed variation $\bbvar{D}$ \cite{Gomes_2016,Kabel_2023}, which is defined by removing all variations that can be absorbed into  a change of the reference fields $X^\mu$ or $\Lambda^\alpha{}_\mu$:
\begin{align}
\bbvar{D}e^\alpha &= \bbvar{d}e^\alpha - \mathcal{L}_{\bbvar{X}}e^\alpha - \ou{\left[\bbvar{D}\Lambda\,\Lambda^{-1}\right]}{\alpha}{\beta}e^\beta\label{eq:e-var}\\
\bbvar{D}A^\alpha_{\;\;\beta}&=\bbvar{d}A^\alpha_{\;\;\beta}-\mathcal{L}_\bbvar{X}A^\alpha_{\;\;\beta}+\nabla\left[\bbvar{D}\Lambda\,\Lambda^{-1}\right]^\alpha_{\;\;\beta},\label{eq:A-var}
\end{align}
where $\bbvar{D}\Lambda$ has an explicit dependence on the coordinate fields. Specifically, we have
\begin{align}
    \bbvar{D}\ou{\Lambda}{\alpha}{\mu}&=\bbvar{d}\ou{\Lambda}{\alpha}{\mu}-\bbvar{X}[\ou{\Lambda}{\alpha}{\mu}]-\ou{[\bbvar{X}\hook A]}{\alpha}{\beta}\ou{\Lambda}{\beta}{\mu}=\bbvar{d}\ou{\Lambda}{\alpha}{\mu}-\nabla_{\bbvar{X}}\ou{\Lambda}{\alpha}{\mu}.\label{eq:Lbda-var}
\end{align}
The covariant field space differentials $\bbvar{D}e^\alpha$ and $\bbvar{D}A^\alpha_{\;\;\beta}$ encode the \emph{physical} variations of the tetrad and connection fields, that is, those variations that cannot be undone by simply performing an appropriate diffeomorphism or internal gauge transformation on the transformed fields.

Inserting \eref{eq:e-var} and \eref{eq:A-var} into \eref{eq:Omdef} and using that $\mathcal{L}_\bbvar{X} A =\bbvar{X}\hook F$, we obtain a bulk and a boundary term for the pre-symplectic form: 
\begin{equation}
    \Omega_\Sigma = \Omega_{\Sigma}^{\text{bulk}} + \Omega_{\partial\Sigma}.\label{eq:Om-blk-bndry-splt}
\end{equation}
The bulk term is
\begin{align}\nonumber
\Omega_{\Sigma}^{\text{bulk}}&=\frac{1}{16\pi G}\int_\Sigma\bigg[\bbvar{D}\Pi_{\alpha\beta}\wedge\bbvar{D}A^{\alpha\beta}+2\bbvar{X}_\alpha\bbvar{D}H^\alpha+(\bbvar{X}\hook G_{\alpha\beta})\wedge\left(\bbvar{D}A+\bbvar{X}\hook F\right)^{\alpha\beta}\\
\nonumber&\qquad-2(\bbvar{D}e_\alpha+\nabla\bbvar{X}_\alpha)\wedge\bbvar{X}\hook H^\alpha-2T_\alpha\bbvar{X}_\beta\wedge\bbvar{X}\hook\left[(\ast-\gamma^{-1})F\right]^{\alpha\beta}\\
&\qquad-G_{\alpha\beta}\left(\bbvar{D}\Lambda\Lambda^{-1}\bbvar{D}\Lambda\Lambda^{-1}\right)^{\alpha\beta}+\left(\bbvar{D}G_{\alpha\beta}+\mathcal{L}_{\bbvar{X}}G_{\alpha\beta}\right)\left(\bbvar{D}\Lambda\Lambda^{-1}\right)^{\alpha\beta}\bigg],\label{eq:Om-blk}
\end{align}
where $\bbvar{X}_\alpha:= \bbvar{X}\hook e_\alpha$ and
\begin{align}
H^\alpha&=\ou{\left[(\ast-\gamma^{-1})F\right]}{\alpha}{\beta}\wedge e^\beta\approx0,\label{eq:Hcons}\\
T_\alpha&=\nabla e_\alpha\approx 0,\label{eq:Tcons}\\
G_{\alpha\beta}&=\nabla\Pi_{\alpha\beta}=2\left(\tfrac{1}{2}\uo{\epsilon}{\alpha\beta}{\alpha'\beta'}-\gamma^{-1}\delta_{[\alpha}^{\alpha'}\delta_{\beta]}^{\beta'}\right)\nabla e_{\alpha'}\wedge e_{\beta'} \approx0\label{eq:Gcons}
\end{align}
denote the constraints. The symbol $\approx$ indicates that the constraints vanish only on-shell, i.e.\ upon imposing the Einstein equations. Note that on-shell, the bulk term reduces to the simple canonical form
\begin{equation}
    \Omega_{\Sigma}^{\text{bulk}}\approx \frac{1}{16\pi G}\int_\Sigma\bbvar{D}\Pi_{\alpha\beta}\wedge\bbvar{D}A^{\alpha\beta}.
\end{equation}
Then, we also have a boundary contribution to the pre-symplectic form. It is
\begin{align}
\Omega_{\partial\Sigma} = \frac{1}{16\pi G}\oint_{\partial\Sigma}\bigg[&\Pi_{\alpha\beta}\left(\bbvar{D}\Lambda\Lambda^{-1}\bbvar{D}\Lambda\Lambda^{-1}\right)^{\alpha\beta}-\left(\bbvar{D}\Pi+\mathcal{L}_{\bbvar{X}}\Pi\right)_{\alpha\beta}(\bbvar{D}\Lambda\Lambda^{-1})^{\alpha\beta} \nonumber \\ 
&+(\bbvar{X}\hook\Pi_{\alpha\beta})\wedge\bbvar{D}A^{\alpha\beta}+\bbvar{X}_\alpha\bbvar{X}_\beta\left((\ast-\gamma^{-1})F\right)^{\alpha\beta}\bigg].\label{eq:Omphysbound}
\end{align}
This boundary contribution  encodes the symplectic structure of the reference fields $X^\mu$ and $\Lambda^\alpha_{\;\;\mu}$ and their conjugate momenta. Since these fields enter the symplectic two-form through a boundary term, they are referred to as \emph{edge} or \emph{boundary modes} (e.g.~\cite{Donnelly_2016, Freidel_2023}) on the covariant phase space. While they encode only gauge-redundant information in the bulk, they become physical at the boundary. That is, to obtain a full description of the subsystem, it is not enough to specify the radiative degrees of freedom alone. In addition, we also need to specify the configuration of the reference fields and their conjugate momenta at the corner $\partial \Sigma$.

\subsection{Linearized Theory}\label{sec:symplecticlinearized}

\noindent Above, we introduced reference frames for covariant gauge fixing conditions using covariant and canonical phase space techniques. The goal of this section is to connect our construction to the more standard procedure in linearized gravity. Thus, we take linearized perturbations around a flat spacetime background $(^{(0)}e^\alpha,^{(0)}\ou{A}{\alpha}{\beta})$. It is important to note that we do not distinguish at this stage between gauge variations and the variations of the true, albeit linearized, physical degrees of freedom. Therefore, we include perturbations of the clock variables around a fiducial background configuration $X_o^\mu$. We thus write
\begin{equation}
X^\mu=X^\mu_o+\delta X^\mu+\mathcal{O}(\delta^2).
\end{equation}
In the same way, we also include perturbations of the internal reference fields $\ou{\Lambda}{\alpha}{\mu}(X)$. Notice that the Lorentz frames carry an explicit dependence on the coordinates. Linearizing both $\Lambda^\alpha{}_\mu$ and $X^\mu$ will  have a combined effect. There are variations $\delta X^\mu$ of the coordinates alone and variations that only change the Lorentz reference frames while keeping $X^\mu$ fixed. We encode variations of $\Lambda(X)$ that do not result from changing $X^\mu$ into an $\mathfrak{so}(1,3)$ Lie algebra element $\ou{\delta\lambda}{\mu}{\nu}$. Combining both variations, we obtain 
\begin{align}\nonumber
\ou{\Lambda}{\alpha}{\mu}(X)&={}^{(0)}\ou{\Lambda}{\alpha}{\mu}(X_o)+{}^{(0)}\ou{\Lambda}{\alpha}{\nu}(X_o)\,\ou{(\delta\lambda)}{\nu}{\mu}\big|_{X_o}+\left(\partial_\nu{}^{(0)}\ou{\Lambda}{\alpha}{\mu}\right)(X_o)\,\delta X^\nu+\mathcal{O}(\delta^2) \\ &\equiv {}^{(0)}\ou{\Lambda}{\alpha}{\mu}(X_o) + {}^{(0)}\ou{\Lambda}{\alpha}{\nu}(X_o)\,\ou{(\delta\barlambda)}{\nu}{\mu}\big|_{X_o},
\end{align}
where, in the second line, we introduced the $\frak{so}(1,3)$-form 
\begin{equation}
\ou{\delta \barlambda}{\nu}{\mu}\equiv\delta\ou{\lambda}{\nu}{\mu}+{}^{(0)}\Lambda_\beta^{\;\;\nu}(X_o)(\partial_\rho {}^{(0)}\Lambda^\beta_{\;\;\mu}(X_o))\delta X^\rho
\end{equation}
to simplify our notation as we move forward. Next, we consider perturbations of the linearized tetrad and connection,
\begin{align}
f^\mu(X)&=(\delta f^\mu)(X_o)+\mathcal{O}(\delta^2),\\
\ou{\Delta}{\mu}{\nu}(X)&=\ou{(\delta\Delta)}{\mu}{\nu}(X_o)+\mathcal{O}(\delta^2).\label{eq:Deltexpans}
\end{align}
The triple $(\delta f^\mu,\delta X^\mu,\delta\ou{\lambda}{\mu}{\nu})$ defines a tangent vector in field space. This tangent vector is based at a flat solution. In turn, we have a perturbative expansion of the tetrads \eref{eq:tetrad} and connection \eref{eq:connection} around this solution. We write
\begin{align}
e^\alpha&={}^{(0)}e^\alpha+\delta e^\alpha+\mathcal{O}(\delta^2)\\
\ou{A}{\alpha}{\beta}&=\tensor[^{(0)}]{A}{^\alpha_\beta}+\delta\ou{A}{\alpha}{\beta}+\mathcal{O}(\delta^2)
\end{align}
where the background configurations are
 \begin{equation}
{}^{(0)}e^\alpha={}^{(0)}\ou{\Lambda}{\alpha}{\mu}(X_o)\di X^\mu_o\\
, \qquad {}^{(0)}\ou{A}{\alpha}{\beta}={}^{(0)}\ou{\Lambda}{\alpha}{\mu}(X_o)(\di{}^{(0)}\uo{\Lambda}{\beta}{\mu})(X_o)\label{eq:Minkwski}.
\end{equation}
Leaving the dependence on $X_o$ implicit\footnote{That is, we write $\delta f^\mu\equiv \delta f^\mu(X_o)$, $\delta \ou{\Delta}{\mu}{\nu}\equiv\delta \ou{\Delta}{\mu}{\nu}(X_o)$, $\delta \ou{\lambda}{\mu}{\nu}\equiv\delta \ou{\lambda}{\mu}{\nu}(X_o)$, ${}^{(0)}\ou{\Lambda}{\alpha}{\mu}\equiv {}^{(0)} \ou{\Lambda}{\alpha}{\mu}(X_o)$, $\ou{\omega}{\mu}{\nu\rho}\equiv\ou{\omega}{\mu}{\nu\rho}(X_o)$.} from now on, the first-order perturbations can be written as
\begin{align}
    \delta e^\alpha &= \ou{\left[{}^{(0)}\Lambda\, (\delta\barlambda){}^{(0)}\Lambda^{-1}\right]}{\alpha}{\beta}{}^{(0)}e^\beta + {}^{(0)}\ou{\Lambda}{\alpha}{\mu}\,(\di\delta X^\mu + \delta f^\mu),\\
    \delta A^\alpha_{\;\;\beta} &= -\ou{\left[{}^{(0)}\Lambda\,\di(\delta\barlambda){}^{(0)}\Lambda^{-1}\right]}{\alpha}{\beta}+\ou{\left[{}^{(0)}
\Lambda(\delta\Delta){}^{(0)}\Lambda^{-1}\right]}{\alpha}{\beta}.
\end{align}

To derive the linearized pre-symplectic form, we need to consider variations of the dynamical fields. In the following, we will keep the background reference field configurations fixed, i.e.
\begin{equation}
\bbvar{d}\tensor[^{(0)}]{\Lambda}{^\alpha_\mu}=0, \qquad \bbvar{d} X^\mu_o=0.
\end{equation}
In this way, the variations of the coordinate fields can only depend on the first-order perturbations and we have
\begin{equation}
\bbvar{X}^\alpha:=\bbvar{X}\hook e^\alpha={}^{(0)}\ou{\Lambda}{\alpha}{\mu}\bbvar{d}({\delta X^\mu})+\mathcal{O}(\delta^2).\label{eq:X-var-linrzd}
\end{equation}
Going back to \eref{eq:e-var}, \eref{eq:A-var} and \eref{eq:Lbda-var}, we then immediately find that
\begin{align}
\bbvar{D}e^\alpha&={}^{(0)}\ou{\Lambda}{\alpha}{\mu}\,\bbvar{d}(\delta f^\mu)+\mathcal{O}(\delta^2),\label{eq:e-var-linrzd}\\
{\bbvar{D}}\ou{A}{\alpha}{\beta}&={}^{(0)}\ou{\Lambda}{\alpha}{\mu}\,\bbvar{d}(\delta \ou{\Delta}{\mu}{\nu}) {}^{(0)}\uo{\Lambda}{\beta}{\nu}+\mathcal{O}(\delta^2),\label{eq:A-var-linrzd}\\
{\bbvar{D}}\ou{\Lambda}{\alpha}{\mu}&={}^{(0)}\ou{\Lambda}{\alpha}{\nu} {\bbvar{d}}\ou{(\delta\barlambda)}{\nu}{\mu}+\mathcal{O}(\delta^2).\label{eq:lmbd-var-linrzd}
\end{align}
Next, we insert the perturbative expansion into the constraints. Going back to \eref{eq:Hcons}, \eref{eq:Tcons}, and \eref{eq:Gcons}, we obtain
\begin{align}
H^\alpha&={}^{(0)}\ou{\Lambda}{\alpha}{\mu}\,\ou{\left[(\ast-\gamma^{-1})\di(\delta\Delta)\right]}{\mu}{\nu}\wedge \di X^\nu_o+\mathcal{O}(\delta^2) \equiv {}^{(0)}\ou{\Lambda}{\alpha}{\mu}\delta H^\mu + \mathcal{O}(\delta^2),\\
T^\alpha&=\,{}^{(0)}\ou{\Lambda}{\alpha}{\mu}\,
\left(\di(\delta f^{\mu})+\ou{[\delta\Delta]}{\mu}{\nu}\wedge\di X^{\nu}_o\right)+\mathcal{O}(\delta^2) \equiv {}^{(0)}\ou{\Lambda}{\alpha}{\mu} \delta T^\mu +  \mathcal{O}(\delta^2),\\
G_{\alpha\beta}&={}^{(0)}\ou{\Lambda}{\alpha'}{\mu}{}^{(0)}\ou{\Lambda}{\beta'}{\nu}\underset{\equiv\delta G^{\mu\nu}}{\underbrace{\left(\epsilon_{\alpha\beta\alpha'\beta'}-2\gamma^{-1}\delta_{\alpha' [\alpha}\delta_{\beta]\beta'}\right)\delta T^\mu\wedge \di X_o^\nu} + \mathcal{O}(\delta^2)}.
\end{align}

Finally, we have all the ingredients at hand to compute the perturbative expansion of the pre-symplectic two-form $\Omega_\Sigma$, which we introduced in the previous section at the non-perturbative level. First, we evaluate $\Omega_\Sigma$ at a configuration, which is flat---that is, we consider the pre-symplectic form for the tangent space $T_{\mathcalvar{p}_o}\mathcal{P}$ around a point $\mathcalvar{p}_o\in \mathcal{P}$ in phase space that describes a flat configuration $\mathcalvar{p}_o=({}^{(0)}e^\alpha, {}^{(0)}\ou{A}{\alpha}{\beta})$ as in \eref{eq:Minkwski} above. We restrict ourselves to a flat solution, while imposing no constraints on the variations $\bbvar{d}e^\alpha$, $\bbvar{d}\ou{A}{\alpha}{\beta}$ of the fundamental fields---thus exploring all directions $T_{\mathcalvar{p}_o}\mathcal{P}$ of tangent space for $\mathcalvar{p}_o$ fixed. We obtain
\begin{align}\nonumber
\Omega_{\Sigma}\Big|_{\mathcalvar{p}_0} &= \frac{1}{16\pi G}\int_\Sigma\bigg[\bbvar{D}\Pi_{\alpha\beta}\wedge\bbvar{D}A^{\alpha\beta}+2\bbvar{X}_\alpha\bbvar{D}H^\alpha+\bbvar{D}G_{\alpha\beta}\left(\bbvar{D}\Lambda\Lambda^{-1}\right)^{\alpha\beta}\bigg]\\
&\quad+ \frac{1}{16\pi G}\oint_{\partial\Sigma}\bigg[\Pi_{\alpha\beta}\left(\bbvar{D}\Lambda\Lambda^{-1}\bbvar{D}\Lambda\Lambda^{-1}\right)^{\alpha\beta}\label{eq:Omlin0}\\ &-\left(\bbvar{D}\Pi+\mathcal{L}_{\bbvar{X}}\Pi\right)_{\alpha\beta}(\bbvar{D}\Lambda\Lambda^{-1})^{\alpha\beta}\nonumber
+(\bbvar{X}\hook\Pi_{\alpha\beta})\wedge\bbvar{D}A^{\alpha\beta}\bigg].\label{eq:Om-T-spaceo}
\end{align}

Next, we insert the perturbative expansion back into \eref{eq:Omlin0}. This way, we obtain the first-order contribution to the pre-symplectic form, ${}^{(1)}\Omega_\Sigma$. We identify three terms:
\begin{equation}
    {}^{(1)}\Omega_\Sigma = {}^{(1)}\Omega_\Sigma^{\text{rad}}+{}^{(1)}\Omega_\Sigma^{\text{diff}} + {}^{(1)}\Omega_\Sigma^{\text{Lorentz}}.\label{eq:omlinearized}
\end{equation}
The first term describes the radiative modes at the linearized level. It consists of the bulk integral
\begin{equation}
{}^{(1)}\Omega_\Sigma^{\text{rad}}=\frac{1}{8\pi G}\int_\Sigma\di X_o^{[\mu}\wedge\bbvar{d}(\delta f^{\nu]}) \big[(\ast-\gamma^{-1})\bbvar{d}(\delta \Delta)\big]_{\mu\nu}.\label{eq:Omlinrzd-rad}
\end{equation}
Next, we have the contribution from the reference fields for the diffeomorphisms. At the linearized level, these are the perturbations $\delta X^\mu$ of the coordinate fields. The resulting contribution to the total pre-symplectic two-form is a sum of two terms:
\begin{equation}
{}^{(1)}\Omega_\Sigma^{\text{diff}}=\frac{1}{8\pi G}\int_\Sigma\bbvar{d}(\delta X^\mu)\,\bbvar{d}(\delta H_\mu)-\oint_{\partial\Sigma}\bbvar{d}(\delta X^\mu)\,\bbvar{d}(\delta \boldsymbol{P}_\mu).\label{eq:diffOmega}
\end{equation}
The first term, which is a bulk integral, vanishes once the constraints are imposed. On-shell, the only non-zero contribution comes from the boundary term, which allows us to identify the \emph{momentum current} $\delta \boldsymbol{P}_\mu$ conjugate to the first-order linearized coordinate fields,
\begin{equation}
 \delta \boldsymbol{P}_\mu=\frac{1}{8\pi G}\big[(\ast-\gamma^{-1})\delta\Delta\big]_{\mu\nu}\wedge\di X^\nu_o.\label{eq:Pcurr}
\end{equation}
From \eref{eq:diffOmega}, we conclude that $X^\mu$ is indeed a reference frame for the constraint, i.e.\  $\{X^\mu(q),H_\nu(q')\}\sim\delta^\mu_\nu\delta^{(3)}(q,q')$ in analogy to \eref{eq:Hxi-X}. This does not imply, however, that the reference frames also Poisson commute among themselves. Instead, we have
\begin{equation}
\big\{\delta X^\mu(q),\delta X^\nu(q')\big\}=\Theta^{\mu\nu}(q,q'),
\end{equation}
for some bi-tensor field $\Theta^{\mu\nu}(q,q')$ that does not vanish on the constraint hypersurface $H_\mu=0$, $G_{\mu\nu}=0$. This non-commutativity has a simple geometric origin. The field space covariant derivative $\bbvar{D}$ introduced in \eref{eq:e-var} and \eref{eq:A-var} removes the gauge variations from the functional derivative of the fundamental fields. It does not remove, however, transversal variations that take us away from the constraint hypersurface $H_\mu=0, G_{\mu\nu}=0$. This has an important consequence. Consider coordinates $(a_\pm,\bar{a}_\pm, H_\mu,X^\mu,\ou{\Lambda}{\alpha}{\mu})$ in a vicinity of the constraint hypersurface such that $(a_\pm,\bar{a}_\pm)$ parametrize the two modes of gravitational radiation thereon.\footnote{In linearized gravity, we can access these modes via a standard projector onto e.g.\ the transverse and traceless modes.} Since $\bbvar{D}$ removes the pure gauge variations from the exterior derivative, we have
\begin{equation}
\bbvar{D}= \int_\Sigma \left[\sum_{s=\pm}\Big(\bbvar{d}a_s\frac{\delta}{\delta a_s}+\bbvar{d}a_s\frac{\delta}{\delta \bar{a}_s}\Big)+\bbvar{d}H_\mu\frac{\delta}{\delta H_\mu}+\bbvar{d}G_{\mu\nu}\frac{\delta}{\delta G_{\mu\nu}}\right].
\end{equation}
If we now return to \eref{eq:Om-blk}, we  see that for any such choice of $(a_\pm,\bar{a}_\pm)$ phase space coordinates the momentum conjugate to $H_\mu$ cannot be the reference frame $X^\mu$ alone. At the linearized level, it is simply $X^\mu$ shifted by terms linear in the constraints. Since the constraints do not commute with the reference fields, this in turn creates the non-commutativity of the embedding fields $X^\mu$. We will comment on this non-commutativity for linearized clocks in Subsection \ref{sec:clockchoices} below.

Finally, there is the symplectic structure for the internal Lorentz reference frames, 
\begin{equation}
{}^{(1)}\Omega_\Sigma^{\text{Lorentz}}=\frac{1}{16\pi G}\int_\Sigma\bbvar{d}(\delta G_{\mu\nu})\,\bbvar{d}(\delta\barlambda)^{\mu\nu}-\frac{1}{2}\oint_{\partial\Sigma}\bbvar{d}(\delta\boldsymbol{S}^{\text{bare}})_{\mu\nu}\,\bbvar{d}(\delta\barlambda)^{\mu\nu}.\label{eq:Om-pertrb-lrntz}
\end{equation}
Again, the first term vanishes upon imposing the constraints and we are left with a boundary term for the first-order linearized Lorentz frames. In here, we introduced the \emph{bare} spin current  $\delta\boldsymbol{S}^{\text{bare}}_{\mu\nu}$, which is conjugate to the internal Lorentz frames. It splits into a background contribution and the contribution for the intrinsic gravitational spin of the perturbation, 
\begin{equation}
\delta\boldsymbol{S}_{\mu\nu}^{\text{bare}}=\delta\boldsymbol{S}_{\mu\nu}^o+\delta\boldsymbol{S}_{\mu\nu},\label{eq:Scurr}
\end{equation}
where
\begin{align}
\delta\boldsymbol{S}_{\mu\nu}^o&=\frac{1}{8\pi G}\left(\tfrac{1}{2}\epsilon_{\mu'\nu'\rho[\mu}-\gamma^{-1}\eta_{\rho[\mu'}\eta_{\nu'][\mu}\right)\di X^{\mu'}_o\wedge\di X^{\nu'}_o\,\ou{\delta\barlambda}{\rho}{\nu]}+\nonumber\\
&\qquad +\frac{1}{4\pi G}\left(\tfrac{1}{2}\epsilon_{\mu\nu\mu'\nu'}-\gamma^{-1}\eta_{\mu[\mu'}\eta_{\nu']\nu}\right)\di X^{\mu'}_o\wedge\di(\delta X^{\nu'}),\\
\delta\boldsymbol{S}_{\mu\nu}&=\frac{1}{4\pi G}\left(\tfrac{1}{2}\epsilon_{\mu\nu\mu'\nu'}-\gamma^{-1}\eta_{\mu[\mu'}\eta_{\nu']\nu}\right)\di X^{\mu'}_o\wedge\delta f^{\nu'}.
\end{align}
We thus find that the covariant phase space for linearized general relativity describes, in addition to gravitational radiation, the linearized reference fields $\delta X^\mu$ for the diffeomorphism group and $(\delta \barlambda)^{\mu\nu}$ for the Lorentz group at the boundary of the spacetime region. Moreover, the explicit form of the symplectic structure \eref{eq:omlinearized} allows us to infer their canonically conjugate momenta: the momentum current $\delta \boldsymbol{P}_\mu$ and the spin current $\delta \boldsymbol{S}_{\mu\nu}$.

\section{Mapping Covariant Reference Frames into the Canonical Phase Space}\label{sec:coordConditions}

\noindent So far, we left the gauge conditions on the reference fields $X^\mu$ and $\ou{\Lambda}{\alpha}{\mu}$ unspecified. The goal of this section is to choose specific gauge conditions and complete the gauge fixing procedure. More specifically, we will see that the covariant gauge fixing conditions introduced in Section \ref{sec:ADMphasespace} indeed reproduce the geometrical clocks, constructed from the linearized Ashtekar-Barbero variables, which are utilized in \cite{Fahn_2022}.\smallskip

\subsection{Choice of clock and spatial reference fields}\label{sec:clockchoices}
The starting point is a perturbative expansion on the extended phase space. As mentioned above, we take the background configuration to be spatially flat with vanishing extrinsic curvature. In terms of standard ADM variables, this amounts to setting
\begin{align}
h_{ab}&={}^{(0)}h_{ab}+\delta h_{ab}+\mathcal{O}(\delta^2)\equiv\delta_{ab}+\delta h_{ab}+\mathcal{O}(\delta^2),\label{eq:ADM-h-prtrb}\\
\tilde{\pi}^{ab}&=\sqrt{\mathrm{det}{}^{(0)}h}\,\delta{\pi}^{ab}+\mathcal{O}(\delta^2)=\delta{\pi}^{ab}+\mathcal{O}(\delta^2),\label{eq:ADM-p-prtrb}
\end{align}
where $\delta_{ab}$ is the flat Euclidean metric and we removed the density weights from the perturbation of the ADM momentum. Notice that the extrinsic curvature vanishes to zeroth order in the expansion. Thus
\begin{equation}
\delta{\pi}^{ab}=\delta K^{ab}-\delta^{ab}\delta K,
\end{equation}
where all indices of the perturbation $(\delta h_{ab}, \delta K_{ab})$ are now raised and lowered with the flat Euclidean reference metric $\delta_{ab}$. \smallskip

Next, we consider the perturbative expansion of the additional configuration variables that we introduced to build the extended phase space. We write
\begin{align}
X^\mu&=X^\mu_o+\delta X^\mu+\mathcal{O}(\delta^2)\\
n^\mu&=\delta^\mu_0+\delta n^\mu+\mathcal{O}(\delta^2),\\
t_\mu&=-\delta_\mu^0+\delta t_\mu+\mathcal{O}(\delta^2).
\end{align}
To leading order in the expansion, the corresponding canonical momenta, vanish i.e.\ $(\tilde{P}_\mu,\tilde{p}_\mu,\tilde{\Pi}^\mu)=\mathcal{O}(\delta)$. The zeroth order of the reference frame $X^\mu$ is adapted to the background metric. We set
\begin{equation}
    X^0 = t + \delta T+\mathcal{O}(\delta^2),\quad
    X^i = x^i + \delta X^i+\mathcal{O}(\delta^2)
\end{equation}
where we choose $X^0_o$ such that $\Sigma$ is a $t=\mathrm{const}.$ surface and $\{x^i\}$ are Cartesian coordinates thereon. In other words,
\begin{equation}
\delta_{ab}=\delta_{ij}\di x^i_a\di x^j_b.
\end{equation}
In the following, $\partial_a$ denotes the corresponding Levi Civita covariant derivative. Since $\{x^i\}$ are Cartesian coordinates, the corresponding Christoffel symbols vanish, i.e.\ $\partial_a\di x^i_b=\partial_a\partial_bx^i=0$. We will need this equation below.

Next, we consider the gauge fixing conditions at the linearized level. First of all, we set
\begin{equation}
\delta n^\mu=0,\qquad \delta t_\mu=0,
\end{equation}
so that the harmonic coordinate condition, namely \eref{eq:gaugecond2} and \eref{eq:gaugecond3} and \eref{eq:NN-def}, simplifies at leading order to
\begin{equation}
    \delta[\Delta_h  X^\mu] - \delta K\delta^\mu_0 +\mathcal{O}(\delta^2)=\delta[\Delta_h  X^\mu] +\frac{1}{2} \delta_{ab}\delta{\pi}^{ab} \delta^\mu_0=0.
\end{equation}
To obtain the first-order harmonic coordinate conditions, we need to take into account the perturbative expansion of the Laplace operator. The spatial metric is perturbed around a flat  background, $h_{ab} = \delta_{ab} + \delta h_{ab}$. We obtain
\begin{align}
    \Delta_h &= h^{ab}D_aD_b = \delta^{ab}\partial_a \partial_b - \delta h^{ab} \partial_a \partial_b - \delta^{ab}\,\delta\ou{\Gamma}{c}{ab}\partial_c+\mathcal{O}(\delta^2),
\end{align}
where $\delta h^{ab} \equiv \delta^{ac}\delta^{bd}\delta h_{cd}$, and $\delta\ou{\Gamma}{c}{ab}$ denotes the perturbation of the Christoffel connection. Using that $\delta^{ab}\ou{\Gamma}{c}{ab}=\partial_a\delta h^{ac} - \frac{1}{2}\partial^c \delta \ou{h}{a}{a}$ with $\delta \ou{h}{a}{a} \equiv \delta^{ac}\delta h_{ac}$. Taking into account $\partial_a\partial_bx^i=0$ and $\partial_a T=0$, we obtain the linearized gauge fixing conditions
\begin{align}
    \Delta(\delta T) &= \delta K = -\frac{1}{2}\delta_{ab}\delta{\pi}^{ab}\label{eq:tcond},\\
    \Delta (\delta X^i) &= \di x^i_b\,\partial_a\left(\delta h^{ab}-\tfrac{1}{2}\delta^{ab}\ou{\delta h}{c}{c}\right),\label{eq:xcond}
\end{align}
where $\Delta=\delta^{ab}\partial_a\partial_b$ is the Laplacian for the flat background metric. 
These equations can be inverted with the Green's function, making the functional dependence between the linearized coordinate fields and the  perturbations of the ADM variables explicit. To see that $(\delta T,\delta X^i)$ are, in fact, the same \emph{geometrical clocks} that were constructed earlier in \cite{Fahn_2022}, we need to switch from the ADM variables $(h_{ab}, \tilde{\pi}^{ab})$ to the Ashtekar--Barbero variables. The Ashtekar--Barbero variables consist of the densitized triad $\uo{\tilde{E}}{i}{a}$ and the Ashtekar--Barbero connection $\ou{A}{i}{a}$. These variables carry an additional $SU(2)$ gauge symmetry, i.e.\ the rotational subgroup of the internal Lorentz symmetry preserving the internal surface normal $n^\alpha=\ou{e}{\alpha}{\mu}n^\mu$ of $\Sigma$. This  internal gauge symmetry is invisible on the ADM phase space. The clock $\delta T$ and spatial reference frame $\{\delta X^i\}$ are invariant under these internal transformations, which is clear given their non-perturbative \eref{eq:gaugecond3} and perturbative definitions \eref{eq:tcond} and \eref{eq:xcond}.  We can thus choose whatever $SU(2)$ gauge we like. To simplify our analysis going forward, we choose a specific $SU(2)$ gauge in which 
\begin{equation}
e^i=\di x^i+\delta e^i+\mathcal{O}(\delta^2).\label{eq:e-prtrb}
\end{equation}
Thus, the background configuration ${}^{(0)}e^i$ is simply the Euclidean coordinate differentials $\di x^i$. Next, we consider the perturbation of the densitized triad \eref{eq:Edens-def} for given co-triadic perturbation $\delta e^i$. First of all, let us recall that it is always possible to rewrite the \emph{densitized triad} $\tilde{E}^{~a}_i$  as 
\begin{equation}
    \tilde{E}^{\;\;a}_i = \mathrm{d}^3x\det(e)\,\uo{e}{i}{a}= \frac{1}{2}\tilde{\epsilon}^{abc}\epsilon_{ijk}e^j_{\;\;b} e^k_{\;\;c},
\end{equation}
where $\epsilon_{ijk}$ denotes the internal Levi-Civita alternating tensor, i.e.\ the structure constants of $\mathfrak{su}(2)$, while $\tilde{\epsilon}^{abc}$ is the Levi-Civita tensor density. It is then immediate to check that the variation of the co-triad can be determined from the variation of the densitized triad through
\begin{equation}
   \di^3 x\,\delta \ou{e}{i}{a}=\left(\frac{1}{2}\di x^i_a\di x^j_b\uo{\delta\tilde{E}}{j}{b}-\di x^i_b\di x^j_a\uo{\delta\tilde{E}}{j}{b}\right).\label{eq:cotriad-var}
\end{equation}
To simplify our notation further, we drop the fiducial density weights $\di ^3x$ from the linear perturbation of $\uo{\tilde{E}}{i}{a}$, thus writing
\begin{equation}
\uo{\tilde{E}}{i}{a} =\di^3x\, \partial^a_i+\di^3x\,\delta\uo{E}{i}{a}+\mathcal{O}(\delta^2).
\end{equation}
\noindent\emph{Choice of linearized clock.} Let us now return to the definition of the linearized clock \eref{eq:tcond} and express it as a functional on the phase space of linearized Ashtekar--Barbero variables. 
The definition \eref{eq:tcond} contains the trace of the extrinsic curvature. In terms of Ashtekar--Barbero variables, the extrinsic curvature is a derived quantity. Using \eref{eq:AC-def}, we have
\begin{equation}
    K^i_{\;\;a} = \frac{1}{\gamma}\left(\ac^i_{\;\;a} - \Gamma^i_{\;\;a}\right).
\end{equation}
The ADM momentum, which is related to  the extrinsic curvature through \eref{eq:AMD-p-def}, vanishes at zeroth order. At this order, the co-triad is simply given by ${}^{(0)}e^i=\di x^i$, see \eref{eq:e-prtrb}. The corresponding $SU(2)$ spin connection coefficients $\Gamma^i_{\;\;a}$ must clearly vanish, and we obtain
\begin{equation}
\ou{A}{i}{a}=\delta \ou{A}{i}{a}+\mathcal{O}(\delta^2).
\end{equation} 
To obtain the variation of the trace of the extrinsic curvature, we note
\begin{equation}
    \delta K = \delta (K^a_{\;\; a}) = \delta (K^i_{\;\;a} e_i^{\;\;a}) 
    = \frac{1}{\gamma}\delta\left(  A^i_{\;\;a} -  \Gamma^i_{\;\;a} \right)\partial^a_i.
\end{equation}
 To proceed further, we simplify the last term in this expression. Note that
\begin{equation}
    \tilde{E}_{i}^{~a}\Gamma^i_{~a} = \frac{1}{2}\tilde{\epsilon}^{abc}\epsilon_{ijk}\ou{\Gamma}{i}{a}e^j_{\;\;b} e^k_{\;\;c} = \frac{1}{2}\ou{\epsilon}{k}{ij}\Gamma^i\wedge e^j\wedge e_k = -\frac{1}{2}\di e^k\wedge e_k,
\end{equation}
where we used in the last step the torsionless condition for the spatial Levi Civita connection on $\Sigma$, i.e.\ $D e^i = \mathrm{d}e^i + \ou{\epsilon}{i}{jk}\Gamma^j\wedge e^k = 0$. This implies further that
\begin{equation}
    \delta(\tilde{E}^{\;\;a}_i \Gamma^i_{\;\;a}) = - \frac{1}{2}\tilde{\epsilon}
    ^{abc}\partial_{a}(\delta e_{kb})\di x^k_c= \frac{1}{2}\ou{\epsilon}{ij}{k}\partial^{a}_i\partial_a(\delta \tilde{E}^{\;\;b}_j)\di x^k_b.
\end{equation}
Thus, the defining equation \eref{eq:tcond} for the linearized time coordinate becomes
\begin{equation}
    \Delta (\delta T) = \frac{1}{\gamma}\left( \partial^a_i \delta A^i_{\;\;a} - \frac{1}{2}\ou{\epsilon}{ij}{k}\partial^a_i\partial_a(\delta E^{\;\;b}_j)\di x^k_b\right).
\end{equation}
We can now solve this gauge condition with the Green's function $G^\Delta$ for the flat Laplace operator $\Delta=\delta^{ab}\partial_a\partial_b$. Our conventions are $\Delta G^\Delta(q,q') = -\delta^{(3)}(x,y)$. We then have
\begin{align}
    (\delta T)(q) 
    &= - \frac{1}{\gamma}\left(  \frac{1}{2}\partial^i_c\epsilon_a^{\;\; cb} \partial_b (\delta E^{\;\;a}_i \ast G^\Delta)(x) + \partial_i^a (\delta A^i_{\;\;a} \ast G^\Delta)(q)\right).\label{eq:dT-clck}
\end{align}
This equation reproduces the geometrical clock that can be found in (2.49) of \cite{Fahn_2022}.

\noindent \emph{Choice of spatial reference frames.} In \eref{eq:xcond}, we defined the spatial reference frames at linear order. To determine them in terms of linearized Ashtekar-Barbero variables, notice first that $\delta h_{ab} = 2 \delta e_{i(a} \di x^i_{b)}=2\delta e_{(ab)}$. Going back to \eref{eq:cotriad-var}, we thus also have
\begin{equation}
\delta h_{ab}=\delta_{ab}\di x^i_c\delta\uo{E}{i}{c}-2\delta E_{(ab)},
\end{equation}
where $\delta E_{ab}=\di x^i_a\delta_{cb}\uo{\delta E}{i}{c}$. This in turn implies with $\delta h^{ab}=\delta^{ac}\delta^{bd}\delta h_{cd}$ that
\begin{equation}
\delta h^{ab}-\frac{1}{2}\delta^{ab}\delta \ou{h}{c}{c}=-2\delta E^{(ab)}+\frac{1}{2}\delta^{ab}\di x^i_c\delta \uo{E}{i}{c}.
\end{equation}
We insert this equation back into  the linearized coordinate condition, i.e. \eref{eq:xcond}. We obtain
\begin{equation}
    \Delta (\delta X^i) = -\di x^i_b\,\partial_a\Big( 2\delta E^{k(b}\partial_k^{a)} - \frac{1}{2}\delta^{ab} \partial^k_c\delta E^{\;\;c}_k \Big).
\end{equation}
Once again, we can solve this equation for $\delta X^i$ as a functional on the phase space of linearized Ashtekar-Barbero variables. Using the Green's function, we obtain 
\begin{equation}
    (\delta X^i)(q)  = \Big( \di x^i_b\partial^{j}_{c} - \frac{1}{2}\di x^j_b\partial^{i}_{c} + \delta^{ij}\delta_{bc} \Big)(\partial^c\delta E^{\;\;b}_j\ast G^\Delta)(q).\label{eq:deltaX-EA}
\end{equation}
This equation reproduces the geometrical clocks introduced in equation (2.53) of \cite{Fahn_2022}. Note that by using the Green’s function $G^\Delta$ to invert the harmonic gauge condition, we are implicitly assuming a specific falloff. Taking into account the asymptotic falloff conditions for the linearized ADM variables,\footnote{Note that the right hand side of \eref{eq:tcond} and \eref{eq:xcond} falls off as $r^{-2}$.} we can see that the reference fields $\delta T$ and $\delta X^i$ can have a finite $r\rightarrow\infty$ limit for generic configurations on phase space. The resulting boundary values $\delta Q^\mu$ depend through \eref{eq:dT-clck} and \eref{eq:deltaX-EA} on the gravitational perturbation $(\delta h_{ab},\delta\pi^{ab})$. Returning to the bulk and boundary symplectic structure \eref{eq:diffOmega}, we can see that the linearized reference fields $\delta X^\mu$ will not drop out completely from the physical phase space. In the bulk, the linearized reference fields are conjugate to the constraints, which vanish on-shell. Hence there is no bulk contribution involving $\delta X^\mu$. At the asymptotic boundary, we are in a different situation. The boundary values $\delta Q^\mu=\lim_{r\rightarrow\infty}\delta X^\mu$, which are functionals on the linearized kinematical ADM phase space, become physical. In other words, they become Dirac observables. Their conjugate variables are the boundary currents given in \eref{eq:Pcurr}. At finite distance, this argument simplifies. At the linearized level, the gauge conditions \eref{eq:tcond} and \eref{eq:xcond} turn into Poisson's equation with a source that depends on the linearized ADM perturbations $(\delta h_{ab},\delta\pi^{ab})$. If we want to solve these equations in a finite domain $\mathcal{D}$, we need to specify e.g.\ Dirichlet boundary conditions for the reference fields, i.e.\ $\delta X^\mu\big|_{\partial \mathcal{D}}=\delta Q^\mu$. The resulting reference fields depend, therefore, on both the boundary conditions $\delta Q^\mu$ and the gravitational perturbation, schematically $\delta X^\mu=\delta X^\mu[\delta h_{ab},\delta\pi^{ab},\delta Q^\mu]$. In the interior of $\mathcal{D}$, the reference fields drop from the symplectic current, because the reference fields are conjugate to constraints that vanish on the physical phase space. On the other hand, there is also a boundary contribution to the symplectic two-form \eref{eq:diffOmega}. This boundary contribution turns the boundary data $\delta Q^\mu$ into elements of the boundary phase space. In this way, boundary data for gauge fixing conditions turn into physical edge modes.\smallskip

A few further remarks: going back to the definition of the linearized clock, i.e.\  \eref{eq:dT-clck} above, we can see that the coordinate frames do not commute among themselves. This is an immediate consequence of the fact that there is a functional dependence on both $\delta\uo{E}{i}{a}$ and $\ou{\delta A}{i}{a}$. This dependence creates a non-commutativity between $\delta T$ and $\Delta X^i$, whereas all spatial reference frames $\delta X^i$ are mutually commuting. In fact,
\begin{equation}
\big\{\delta T(q),\delta X^i(q')\big\}=-4\pi G\di x^i_b\partial^b\Delta^{-2}\delta^{(3)}(q,q').
\end{equation}
This non-commutativity extends to the internal reference frames, which are conjugate to the Gauss constraint. In the linearized theory, this non-commutativity can be always removed by shifting the coordinate functions by an auxiliary contribution which is proportional to the constraints, see \cite{Fahn_2022} for a systematic way to add such terms through a dual observable map to obtain mutually commuting geometrical clocks. This method can be applied order by order in perturbation theory and hence can be used beyond the linearized theory. Here we observe this non-commutativity directly at the level of the geometric clocks. 

An alternative way to obtain this result is to calculate the Dirac bracket associated with the gauge fixings for $\delta n^\mu, \delta t_\mu$ and $\delta\Phi^\mu$. Since $\delta X^\mu$ and $\delta \tilde{P}_\mu$ both commute with $\delta n^\mu, \delta t_\mu$ and their conjugate momenta, the Dirac bracket associated with these gauge fixings for $\delta X^\mu$ and $\delta\tilde{P}_\mu$ coincides with the ordinary Poisson bracket. However, this is no longer the case if we consider the second-class pair $(\delta \Phi^\mu,\delta {\bf H}^\mu)$. Neither $\delta X^\mu$ nor $\delta \tilde{P}^\mu$ commute with $\delta {\bf H}^\mu$ and, moreover, $\delta\tilde{P}_\mu$ does not commute with $\delta X^\mu$. Consequently, there are non-trivial additional contributions in the Dirac bracket compared to the Poisson bracket, leading to a modified algebra for $\delta X^\mu$ and $\delta\tilde{P}_\mu$. In particular, we have $\{\delta X^\mu,\delta X^\nu\}\not=0$  and expressing this result in terms of Ashtekar-Barbero variables results in the non-commutativity discussed above.

\subsection{Choice of Internal Lorentz Frames}
\label{sec:LorntzChoice}

\noindent After having introduced a linearized time function $\delta T$ and spatial reference frames $\delta X^i$, we are left to choose a reference frame for the internal Lorentz symmetries. This amounts to a mere gauge fixing for the spin connection, or, equivalently, a choice of an orthonormal basis $\boldsymbol{E}^\alpha$ in the Lorentz bundle. Going back to \eref{eq:tetrad}, we first identify the reference frame $\boldsymbol{E}^\alpha$ with the $SO(1,3)$ gauge elements $\ou{\Lambda}{\alpha}{\mu}$. More specifically, we write 
\begin{align}
\boldsymbol{E}^\alpha&=(\ou{\Lambda}{\alpha}{0},\ou{\Lambda}{\alpha}{1},\ou{\Lambda}{\alpha}{2},\ou{\Lambda}{\alpha}{3}).
\end{align}
In here and below, boldface characters are columns, rows or matrices. The dual basis of $\boldsymbol{E}^\alpha$ is
\begin{equation}
\boldsymbol{e}_\alpha=\left(\begin{smallmatrix}-\Lambda_{\alpha0}\\\phantom{+}\Lambda_{\alpha1}\\\phantom{+}\Lambda_{\alpha2}\\\phantom{+}\Lambda_{\alpha3}
\end{smallmatrix}\right)
\end{equation}
The two bases satisfy
\begin{equation}
\eta_{\alpha\beta}\boldsymbol{E}^\alpha\otimes\boldsymbol{E}^\beta =\operatorname{diag}(-1,+1,+1,+1),\quad\boldsymbol{e}_\alpha\otimes\boldsymbol{E}^\alpha=\operatorname{id}.\label{eq:orthnorm}
\end{equation}
One possible choice is given by Lorentz gauge, which is the same as to say\footnote{The seemingly simpler gauge condition
$\square_g\boldsymbol{E}^\alpha=g^{\mu\nu}\nabla_\mu\nabla_\nu\boldsymbol{E}^\alpha=0$
is inconsistent with the normalization \eref{eq:orthnorm}.}
\begin{equation}
g^{\mu\nu}\nabla_\mu\left[\boldsymbol{e}_\alpha\otimes\nabla_\nu\boldsymbol{E}^\alpha\right]\equiv\nabla_\mu\boldsymbol{\Delta}^\mu=0,\label{eq:Lorntzgauge}
\end{equation}
where $\boldsymbol{\Delta}_\mu$ is an $\mathfrak{so}(1,3)$-valued one-form, whose entries are given in \eref{eq:connection}, and $\nabla_\mu$ is the covariant derivative. It acts on internal Lorentz indices $\alpha,\beta,\dots$ through the $SO(1,3)$ spin connection and on space-time indices through the Christoffel symbols, but does not mix the columns of $\boldsymbol{E}^\alpha$, i.e.\ $\nabla\ou{\Lambda}{\alpha}{\mu}=\di \ou{\Lambda}{\alpha}{\mu}+\ou{A}{\alpha}{\beta}\ou{\Lambda}{\beta}{\mu}$.\smallskip

While \eref{eq:Lorntzgauge} seems like a natural gauge condition from a Lagrangian point of view, it is less so in the Hamiltonian framework, when working with $SU(2)$ Ashtekar--Barbero variables. In that case, a better choice was introduced in \cite{Fahn_2022}. It is easy to check, in fact, that the linearization of the gauge conditions \eref{eq:Lorntzgauge} does not reproduce the internal reference frames of \cite{Fahn_2022}. This is so because the internal reference frames of \cite{Fahn_2022} are constructed out of the linearized Ashtekar--Barbero connection, whereas \eref{eq:Lorntzgauge} is a functional of only the Lorentz connection.\smallskip 

To translate the perturbative reference frames $\delta \Xi^i(x)$ of \cite{Fahn_2022} into our framework, we proceed in two steps. First, we lift the $SU(2)$ Ashtekar--Barbero connection, which is otherwise only defined on a spatial hypersurface, into spacetime. Then, we replace in \eref{eq:Lorntzgauge} the $SO(1,3)$ covariant derivative $\nabla_\mu$ by the four-dimensional lift of the $SU(2)$ Ashtekar-Barbero covariant derivative. This construction requires a few additional building blocks. First of all, we introduce an $3+1$-split in internal space \cite{Freidel_2020_em2, Wieland_2010}. We set
\begin{equation} 
h_{\alpha\beta}:=\eta_{\alpha\beta}+n_\alpha n_\beta,\quad\text{and}\quad\epsilon_{\alpha\beta\gamma}:=\epsilon_{\delta\alpha\beta\gamma}n^\delta.
\end{equation}
For any Lorentz vector valued $p$-form $V^\alpha\in\Omega^p(\mathcal{M}:T\mathbb{M})$, we can now introduce the covariant Ashtekar--Barbero exterior derivative, namely
\begin{equation}
\mathcal{D}V^\alpha=\nabla V^\alpha-(n^\alpha\nabla n_\beta-n_\beta\nabla n^\alpha)\wedge V^\beta+\gamma\ou{\epsilon}{\alpha}{\delta\beta}\nabla n^\delta\wedge V^\beta.
\end{equation}
This connection has the following key properties
\begin{align}
\mathcal{D}\eta_{\alpha\beta}=0,\qquad
\mathcal{D}n^\alpha=0.
\end{align}
Notice also that the pullback of $\mathcal{D}$ to $\Sigma$ defines the standard $SU(2)$ Ashtekar-Barbero connection for a given foliation if we identify $n^\alpha\uo{e}{\alpha}{a}$ with the normal vector field orthogonal to each leaf. This is often referred to as \emph{time gauge} in the literature, see e.g.\ \cite{Alexandrov_2003,Freidel_2020_em3,Wieland_2010}. It is the choice which we also impose here.\smallskip

The relationship between $\mathcal{D}$ and the $SU(2)$ Ashtekar-Barbero connection can be made more explicit by introducing the $3+1$ split of the reference fields $\boldsymbol{E}^\alpha$. We set
\begin{equation}
\boldsymbol{E}^\alpha=(\ou{\Lambda}{\alpha}{0},\ou{\Lambda}{\alpha}{i})=(n^\alpha,\ou{\lambda}{\alpha}{i}).
\end{equation}
From \eref{eq:orthnorm}, it follows that
\begin{equation}
n_\alpha\ou{\lambda}{\alpha}{i}=0,\qquad \eta_{\alpha\beta}\ou{\lambda}{\alpha}{i}\ou{\lambda}{\beta}{j}=\delta_{ij}.
\end{equation}
The tensor field $\ou{\lambda}{\alpha}{i}$ provides a soldering form (a sort of projector) between the Lorentz bundle and the internal $SU(2)$ frame bundle. If, for example, $\varphi:\Sigma\hookrightarrow\mathcal{M}$ is the embedding of the spatial hypersurface into spacetime, the soldering forms $\ou{\lambda}{\alpha}{i}$ allow us to introduce the standard triads $e^i$ and the extrinsic curvature $K^i$ simply by
\begin{align}
e^i=\varphi^\ast(\uo{\lambda}{\alpha}{i}e^\alpha),\qquad
K^i=\varphi^\ast(\uo{\lambda}{\alpha}{i}\nabla n^\alpha),
\end{align}
where all indices are raised and lowered using the flat internal metrics $\eta_{\alpha\beta}, \delta^{ij}$. The $SU(2)$ spin connection $\Gamma^i$ is then the unique solution of 
\begin{equation}
    \di e^i+\ou{\epsilon}{i}{jk}\Gamma^j\wedge e^k=0,
\end{equation}
where $\epsilon_{ijk}=\epsilon_{\alpha\beta\gamma}\ou{\lambda}{\alpha}{i}\ou{\lambda}{\beta}{j}\ou{\lambda}{\gamma}{k}$.\smallskip

Sine $\mathcal{D}$ annihilates $n^\alpha$ and since $\boldsymbol{E}^\alpha=(n^\alpha,\ou{\lambda}{\alpha}{i})$ is a basis in the frame bundle, all non-vanishing connection coefficients are determined by $\mathcal{D}$ acting on $\ou{\lambda}{\alpha}{i}$. These in turn assume the standard form of the $SU(2)$ Ashtekar-Barbero connection coefficients, which are the sum of the components of the intrinsic $SU(2)$ spin connection and the extrinsic curvature. We obtain
 \begin{equation}
    \mathcal{D}_a\ou{\lambda}{\alpha}{j}=\ou{\epsilon}{i}{kj}\ou{A}{k}{a}\ou{\lambda}{\alpha}{i}=\ou{\epsilon}{i}{kj}\left(\ou{\Gamma}{k}{a}+\gamma\ou{K}{k}{a}\right)\ou{\lambda}{\alpha}{i},\label{eq:LorentAshtekargauge}
\end{equation}

We are now ready to introduce the gauge condition for the $SU(2)$ Ashtekar-Barbero connection. It reads
\begin{equation}
g^{\mu\nu}\mathcal{D}_\mu\left[\uo{\lambda}{\alpha}{i}\mathcal{D}_\nu\ou{\lambda}{\alpha}{j}\right]=g^{\mu\nu}\partial_\mu\left[\uo{\lambda}{\alpha}{i}\mathcal{D}_\nu\ou{\lambda}{\alpha}{j}\right]=0.
\end{equation}

Finally, let us compare this gauge condition with the linearized reference frames introduced in \cite{Fahn_2022}.
Now working in a perturbative framework, we expand the soldering forms $\ou{\lambda}{\alpha}{i}$ into a background plus perturbation,
\begin{equation}
\ou{\lambda}{\alpha}{i}={}^o\ou{\lambda}{\alpha}{i}+\delta\ou{\lambda}{\alpha}{i}
\end{equation}
Following \cite{Fahn_2022}, we assume the background to be flat. More specifically, we take it to be flat with respect to both the covariant Ashtekar-Barbero connection as well as the Lorentz connection itself, i.e.\
\begin{equation}
{}^o\mathcal{D}\,{}^o\ou{\lambda}{\alpha}{i}={}^o\nabla{}\ou{\lambda}{\alpha}{i}=0,
\end{equation}
where ${}^o\mathcal{D}$ and ${}^o\nabla$ denotes the zeroth order of the perturbative expansion of the respective derivatives.
Next, we consider these reference frames at the linearized level. We introduce the perturbed reference frame $\delta \Xi^i$ through the definition
\begin{equation}
\delta\Xi^i:=\frac{1}{8\pi G}\uo{\epsilon}{j}{ik}\,{}^o\uo{\lambda}{\alpha}{j}\delta\ou{\lambda}{\alpha}{k}.\label{eq:Xi-gauge-cond-lin}
\end{equation}

It is now  straightforward to show that to leading order into the expansion, the gauge condition \eref{eq:LorentAshtekargauge} translates into
\begin{equation}
{}^o\square \delta \Xi^i=-\frac{1}{4\pi G}\partial_\mu\delta{A}^{i\mu},
\end{equation}
which matches the definition of the linearized reference frames of \cite{Fahn_2022} if we set in here all time derivatives to zero. In this way, we obtain
\begin{equation}\label{eq:Xi-gauge-cond-lin2}
\delta \Xi^i(x) = \frac{1}{4\pi G}\partial^a (\delta A_a^{\;\; i}\ast G^\Delta)(x).
\end{equation}

\section{Relation of Our Results to Other Work}\label{sec:relationOtherWork}

\noindent We conclude with a general discussion on how our results connect to related developments in the area. First, we comment on other possible  extensions of the ADM phase space (Subsection \ref{subsec:relADMphaseSpace}). Second, we explain the effects of the Barbero--Immirzi parameter on the boundary currents and charges in the covariant phase space formalism (Subsection \ref{sec:boundaryCurrents}). Finally, we speculate about potential consequences for the quantum theory, in particular on higher multipoles and quantum reference frame transformations (Subsection \ref{sec:BIparameter}).

\subsection{Relation to existing work in ADM Phase Space}\label{subsec:relADMphaseSpace}
\noindent 
In the previous subsection, we established an extended ADM phase space that contains reference frames $X^\mu:\Sigma\rightarrow\mathbb{M}$. These reference frames form a second-class constraint pair with the four-dimensional diffeomorphisms generators $\boldsymbol{H}^\mu$ given in \eqref{eq:DefHmu}. To compare our construction to earlier results developed in \cite{Pons_1996,Pons_1999,Garcia_2000,Pons_2003,Pons_2009,Pons_2010,Giesel_2017,Giesel_2018}, where a different extension of the reduced ADM phase space is given, in which the lapse function and shift vector are promoted into elementary canonical variables, we examine here the Hamiltonian generators of diffeomorphisms further and compare them to the Lagrangian theory. We then use our results to discuss the similarities and differences to the extended phase space given in \cite{Pons_1996,Pons_1999,Garcia_2000,Pons_2003} and  \cite{Isham_1984b,Isham_1984a, Kouletsis_2008}. Finally, we will  compare our formalism to earlier results on this topic at the canonical level, see \cite{Kuchar_1991}.

The starting point is the extended phase space introduced in Subsection \ref{sec:ExtADM}. Given a point $(h_{ab},X^\mu,n^\mu,t_\mu;$ $\tilde{\pi}^{ab},$ $\tilde{P}_\mu,\tilde{\Pi}^\mu,\tilde{\pi}_\mu)$ on the extended phase space, we have initial data to integrate the Hamiltonian flow of the extended canonical Hamiltonian $\boldsymbol{H}$, see \eqref{eq:Hdef}. Besides the usual ADM initial data, which consists of configurations of the three-metric $h_{ab}$ intrinsic to $\Sigma$ and the ADM conjugate momentum $\tilde{\pi}^{ab}$, there is in addition auxiliary initial data for the embedding fields $X^\mu:\Sigma\rightarrow\mathbb{M}$ on $\Sigma$. To integrate Hamilton's equations, we then also need to choose a configuration of $n^\mu:\mathbb{M}\rightarrow T\mathbb{M}$ and $t_\mu:\mathbb{M}\rightarrow T^\ast\mathbb{M}$, which, along with their conjugate momenta, are both part of phase space as well. The initial data given by any such $(t^\mu(x),n_\mu(x))$ is rather unusual. In field theory, phase space is usually built from fields at co-dimension one boundaries. Here, we are in a different situation: both $n^\mu(x)$ and $t_\mu(x)$ have a four-dimensional functional dependence on points $x\in\mathbb{M}$ in target space. This is unusual, but we are led to this specific choice to accommodate for the gauge modes that the spacetime metric and its conjugate momenta inherit from the Lagrangian theory. 
A different choice would be to follow the construction given in \cite{Pons_1996,Pons_1999,Garcia_2000,Pons_2003}. This approach is, however, less useful for our present goal to promote the harmonic gauge condition of the Lagrangian theory into a gauge-fixing constraint on a suitably extended canonical phase space.  

As discussed above, the additional gauge modes are removed by a total of 16 constraints.  The conjugate momenta $\tilde{P}_\mu$, $\tilde{\Pi}^\mu$ and $\tilde{\pi}_\mu$ to $X^\mu$, $t_\mu$ and $n^\mu$ vanish as primary constraints. Moreover, the Hamiltonian flow of the extended canonical Hamiltonian preserves the primary constraints provided the ADM initial data $(h_{ab},\tilde{\pi}^{ab})$ satisfies the Hamiltonian and spatial diffeomorphism constraints. Given the extended canonical Hamiltonian \eqref{eq:Hdef}, we can now generate a one-parameter family of fields $(h_{ab}^{(t)},\tilde{\pi}^{ab}_{(t)},X^\mu_{(t)})_{t\in\R}$ intrinsic to $\Sigma$, such that
\begin{align}
\frac{\di}{\di t} h_{ab}^{(t)}&=\left\{h_{ab}^{(t)},\boldsymbol{H}\right\}\equiv\mathfrak{X}_{\boldsymbol{H}}[h_{ab}^{(t)}],\\
\frac{\di}{\di t} \tilde{\pi}^{ab}_{(t)}&=\left\{\tilde{\pi}^{ab}_{(t)},\boldsymbol{H}\right\}\equiv\mathfrak{X}_{\boldsymbol{H}}[\tilde{\pi}^{ab}_{(t)}],\\
\frac{\di}{\di t} X^\mu_{(t)}&=\left\{X^\mu_{(t)},\boldsymbol{H}\right\}\equiv\mathfrak{X}_{\boldsymbol{H}}[X^\mu_{(t)}],
\end{align}
where $\mathfrak{X}_{\boldsymbol{H}}[\cdot]=\{\cdot,\boldsymbol{H}\}$ is the Hamiltonian vector field of the extended canonical Hamiltonian $\boldsymbol{H}$ on the extended phase space.
We proceed to build the metric tensor on the spacetime manifold $\mathcal{M}=\R\times\Sigma$. First of all, we define an auxiliary time coordinate $t$ such that $t(\tau,q)=\tau$ for all $(\tau,q)\in\mathcal{M}=\R\times\Sigma$. In \eref{eq:mtrc-def}, we introduced the coefficients of the inverse spacetime metric  $g^{\mu\nu}$ restricted to $\Sigma$  with respect to the embedding fields $X^\mu$. Given initial data,  the Hamiltonian flow can now generate an entire $t$-parameter family $\{g^{\mu\nu}_{(t)}\}$ of such four-metrics on $\Sigma$. If the inverse metric $g^{\mu\nu}_{(t)}$ is invertible for all $t$, we obtain a four-dimensional metric tensor $g_{\mu\nu}$ as a $t$-parameter trajectory on the extended phase space.\smallskip

In \eqref{eq:Hxi-def}, we further introduced a Hamiltonian generator $\boldsymbol{H}_\xi$, which is related to four-dimensional diffeomorphisms. In fact, it is immediate to see that the Hamiltonian vector field  $\mathfrak{X}_{\boldsymbol{H}_\xi}=\{\cdot,\boldsymbol{H}_\xi\}$ generates the usual ADM representation of four-dimensional diffeomorphisms in terms of $h_{ab}$ and $\tilde{\pi}^{ab}$. In addition, it also generates shifts of $X^\mu$, see  \eqref{eq:Hxi-X}.  If, however, we act with the Hamiltonian vector field on $g_{\mu\nu}$ itself, there is a deviation, i.e.\ $\mathfrak{X}_{\boldsymbol{H}_\xi}[g_{\mu\nu}]\neq \mathcal{L}_\xi g_{\mu\nu}$. The reason for this deviation is that, while the generator $\boldsymbol{H}_\xi$ in \eqref{eq:Hxi-X} is a functional of $h_{ab},\tilde{\pi}^{ab}, X^\mu,\tilde{P}_\mu$ and $n^\mu$, it does not depend on $t_\mu,\tilde{p}^\mu$ and $\widetilde{\Pi}^\mu$. This implies that the action of the Hamiltonian vector field $\mathfrak{X}_{\boldsymbol{H}_\xi}$ on, in particular, $n^\mu$ is trivial and does not describe diffeomorphisms. To obtain a generator that acts as a four-dimensional diffeomorphism on the spacetime  metric coefficients $g_{\mu\nu}$, we need to extend the generator $\boldsymbol{H}_\xi$ so that it has the correct action on the composite functions that define the lapse function and shift vector and their conjugate momenta, see e.g.\ \eref{eq:NN-def} above.

For this purpose, we identify appropriate coordinates on the extended phase space that are conjugate to lapse and shift as introduced in \eqref{eq:NN-def}. We define the following quantities:
\begin{align}
{}^{(3)}\tilde{\Pi}^\mu&:=\int_{-\infty}^\infty\di t\,\{\boldsymbol{H},X^\nu_{(t)}\}\big(\partial_\nu\hook\tilde{\Pi}^\mu\big)(X_{(t)}),\label{eq:three-NPi}\\
{}^{(3)}\tilde{p}_\mu&:=\int_{-\infty}^\infty\di t\,\{\boldsymbol{H},X^\nu_{(t)}\}(\partial_\nu\hook\tilde{p}_\mu)(X_{(t)}).\label{eq:three-Np}
\end{align}
where $X^\mu_{(t)}=\exp(t\mathfrak{X}_{\boldsymbol{H}})^\ast X^\mu$ is the pull-back on phase space of $X^\mu$ under the Hamiltonian flow generated by the extended canonical $\boldsymbol{H}$. Note that $\tilde{\Pi}^\mu$ and $\tilde{\pi}_\mu$ are (co)vector-valued four-densities on target space, in other words, four-forms. In addition, $\partial_\mu\hook\cdot$ denotes the interior product between these four-forms and the tangent vectors $\frac{\partial}{\partial X^\mu}$ on target space. Upon performing the one-dimensional smearing,  ${}^{(3)}\tilde{\Pi}^\mu$ and ${}^{(3)}\tilde{\Pi}_\mu$  are thus (co)vector-valued three-densities on $\Sigma$.
It is now easy to check that the fundamental Poisson brackets \eqref{eq:Poiss1}, \eqref{eq:Poiss2}, \eqref{eq:Poiss3} imply
\begin{align}
\big\{(t_\mu\circ X)(q),{}^{(3)}\tilde{\Pi}^\nu(q')\big\}&\approx\delta^\nu_\mu\tilde{\delta}_\Sigma^{(3)}(q,q'),\label{eq:newPi1}\\
\big\{(n^\mu\circ X)(q),{}^{(3)}\tilde{p}_\nu(q')\big\}&\approx\delta_\nu^\mu\tilde{\delta}_\Sigma^{(3)}(q,q').\label{eq:newPi2}
\end{align}
Thus, the spatial and normal components of ${}^{(3)}\tilde{\Pi}^\mu$ and ${}^{(3)}\tilde{\pi}_\mu$ are each conjugate to lapse and shift in \eqref{eq:NN-def}, e.g.\ $\{N(q),\tilde{\Pi}^\mu(q)(n_\mu\circ X)(q')\}=\tilde{\delta}^{(3)}_\Sigma(q,q')$. Furthermore, the momenta each vanish as a primary constraints, i.e.\ ${}^{(3)}\tilde{\Pi}^\mu\approx 0$ and ${}^{(3)}\tilde{\pi}_\mu\approx 0$.

As expected, these canonical momenta are defined through a non-local smearing with respect to the target space time coordinate. This is different from the situation in \cite{Pons_2003,Giesel_2017,Giesel_2018}, where the conjugate momenta of lapse and shift are still local quantities. As discussed in the previous subsection, this is due to our choice of elementary variables in the extended phase space. This choice was mainly guided by our aim to implement the covariant harmonic gauge fixing condition as a local constraint on the extended phase space. 

After having defined the smeared momenta \eref{eq:newPi1} and \eref{eq:newPi2}, we can now introduce the improved generator
\begin{equation}
\boldsymbol{G}_\xi = \boldsymbol{H}_\xi+\int_\Sigma[t,\xi]_\mu{}^{(3)}\tilde{\Pi}^\mu-\int_\Sigma(\mathcal{L}_\xi n_\nu)\partial^aX^\nu \partial_a X^\mu{}^{(3)}\tilde{p}_\mu.\label{eq:Gdef}
\end{equation}
In the following, we view $\ou{\di X}{\mu}{a}:=\partial_aX^\mu$ as an abstract soldering form between $T\Sigma$ and $T\mathbb{M}$, with all its indices raised and lowered with respect to the respective metric tensors, e.g.\ $\uo{\di X}{\mu}{a}=h^{ab}g_{\mu\nu}\partial_bX^\nu$. Now, the second term contains the projection of the Lie derivative $h^{\mu\nu}\mathcal{L}_\xi n_\nu$  of the co-normal $n_\mu$. This term can be expressed implicitly as a functional on the extended phase space. A short calculation gives
\begin{equation}
\di X^{\mu a}\mathcal{L}_\xi n_\mu=\xi_\perp\beta^a-\partial^a\xi_\perp,
\end{equation}
where $\beta^a$ is the acceleration, see \eref{eq:acclrtn}. Both additional terms involve primary constraints so that $\boldsymbol{H}_\xi$ and $\boldsymbol{G}_\xi$ are weakly equivalent. The first term, on the other hand, contains the $T\mathbb{M}$-valued Lie bracket $[t,\xi]^\mu$, which is defined on phase space by
\begin{equation}
[t,\xi]^\mu:=\left\{\{X^\mu,\boldsymbol{H}_\xi\},\boldsymbol{H}\right\}-\left\{\{X^\mu,\boldsymbol{H},X^\mu\},\boldsymbol{H}_\xi\right\}=\left\{X^\mu,\{\boldsymbol{H}_\xi,\boldsymbol{H}\}\right\}.
\end{equation}
Furthermore, target space indices are raised and lowered using the metric tensor $g^{\mu\nu}$ (and $g_{\mu\nu}$), which is defined as a functional on the extended phase space in \eref{eq:mtrc-def} above. In other words, $[t,\xi]_\mu=[t,\xi]^\nu g_{\nu\mu}$.\smallskip

To conclude this subsection, we compare our results to those of \cite{Lee_1990,Pons_1996,Pons_1999,Garcia_2000,Pons_2003,Pons_2009,Pons_2010}.
First of all, since $(h_{ab},\tilde{\pi}^{ab})$ commute with the extension of $\boldsymbol{H}_\xi$ in the improved generators $\boldsymbol{G}_\xi$, the Hamiltonian vector field of $\boldsymbol{G}_\xi$ reproduces the standard ADM commutation relations. To see this explicitly, we introduce the decomposition of the target space tangent vector $\xi^\mu(X)$ into its spatial and temporal components on $\Sigma$. This reads
\begin{equation}
\xi^\perp=-n^\mu(X)\xi_\mu(X),\qquad \xi_\parall^a =\di X^{\mu a}\xi_\mu(X).
\end{equation}
Then we obtain
\begin{equation}
\{h_{ab},\boldsymbol{G}_\xi\}=2\left(\xi^\perp K_{ab}+D_{(a}\xi^\parall_{b)}\right),
\end{equation}
Next, we compute the action of the improved diffeomorphism generator $\boldsymbol{G}_\xi$ on the lapse function $N$. As shown in Appendix \ref{app:CompPons}, we get
\begin{align}\label{eq:GxiN1}
\big\{N,\boldsymbol{G}_\xi\big\}&=
-Nn^\mu\nabla_\mu\xi^\nu n_\nu=Nn^\mu\nabla_\mu\xi_\perp+N\xi_\parall^a\beta_a.
\end{align}
Taking into account that $t^\mu=N n^\mu+\ou{\di X}{\mu}{a} N^a$ and the definition of the acceleration vector $\beta^a$, see \eqref{eq:acclrtn}, the above result can be rewritten as
\begin{equation}
\big\{N,\boldsymbol{G}_\xi\big\}=\dot{\xi}_\perp-N^a D_a\xi_\perp+\xi^a_\parall D_a N,\label{eq:GxiN2}
\end{equation}
where $\dot{\xi}_\perp=-t^\nu\partial_\nu(n^\mu \xi_\mu)=\{\xi_\perp,\boldsymbol{H}\}$. This result agrees exactly with the result in \cite{Pons_1996,Giesel_2017,Giesel_2018}. However, there is a difference in the treatment of the descriptor $\xi$. While its temporal derivative $\dot{\xi}_\perp$ can be expressed on the extended ADM phase space in our approach, the descriptor $\xi$, as well as its orthogonal and spatial projections, are external parameters in \cite{Pons_1996,Pons_1999,Garcia_2000,Pons_2003,Giesel_2017,Giesel_2018}.

Finally, we consider how the improved diffeomorphism generator acts on the shift vector. Going back to the definition of \eqref{eq:Gdef}, we end up with
\begin{align}
\nonumber \ou{\di X}{\mu}{a}\big\{N^a,\boldsymbol{G}_\xi\big\}&=\ou{h}{\mu}{\nu}[Nn,\xi_\perp n+\xi_\parall]^\nu+N \di X^{\mu a}(\xi_\perp\beta_a-D_a\xi_\perp).
\end{align}
Now, since $Nn^\mu=t^\mu-\ou{\di X}{\mu}{a} N^a$, we obtain
\begin{equation}
\big\{N^a,\boldsymbol{G}_\xi\big\}=\dot{\xi}^a_\parall-N^bD_b\xi^a_\parall+\xi^b_\parall D_bN^a+h^{ab}(\xi_\perp D_bN-ND_b\xi_\perp),
\end{equation}
where 
\begin{equation}
\dot{\xi}^a_\parall=\big\{\di X^{\mu a}\xi_\mu(X),\boldsymbol{H}\big\}=\uo{\di X}{\mu}{a}\mathcal{L}_t(\ou{\di X}{\mu}{b}\xi^b_\parall).
\end{equation}
Here, too, we obtain agreement with the results in \cite{Pons_2003}. Just as in the case of the lapse function, we realize the temporal derivative $\dot{\xi}^a_\parall$ on the extended ADM phase space, which  contrasts earlier work on the topic, see \cite{Pons_2003,Giesel_2017,Giesel_2018}.\smallskip

In summary, the extended phase space proposed in this article is consistent with the extension given in \cite{Pons_1996,Pons_1999,Garcia_2000,Pons_2003}. However, there are also differences. The first difference is that the extended ADM phase space introduced in here is slightly larger. It also contains the embedding fields $X^\mu$ themselves. Second, the underlying field content is different in the two extensions. While in \cite{Pons_1996,Pons_1999,Garcia_2000,Pons_2003}, lapse and shift and their momenta are added as canonical variables, here, in addition to the embedding fields, the extended phase space also involves $t_\mu$ and $n^\mu$ together with their conjugate momenta. Consequently, lapse and shift are expressed as composite functions of the additional phase space variables, see \eref{eq:NN-def} above. The same applies to their momenta, which can be reconstructed as temporally non-local functions, see \eref{eq:three-NPi} and \eref{eq:three-Np}. \smallskip

For the aim of linking geometrical clocks with edge modes, it was pivotal, that there is a canonical generator for four-dimensional diffeomorphisms that has the correct action on all metric components, not only the ADM data, but also on lapse and shift with the additional advantage that the embedding fields $X^\mu$ are part of phase space as well. This is provided in the extended phase space in \cite{Pons_1996,Pons_1999,Garcia_2000,Pons_2003} a property also shared in our formalism.
Another crucial observation is that temporal derivatives of the descriptor $\xi$ of the improved generator $\boldsymbol{G}_\xi$ can be expressed entirely in terms of the extended ADM phase space introduced in Subsection \ref{sec:ExtADM}. This has the advantage that successive applications of $\boldsymbol{G}_\xi$ also directly correspond to the action of four-dimensional diffeomorphisms in the Lagrangian formulation, since the involved time derivative of $\xi$ is automatically taken into account in expressions such as ${\cal L}_\xi({\cal L}_\xi(g))_{\mu\nu}$.\smallskip

Finally, let us compare the extended phase space introduced here, with other works where embedding fields are included in an extension of the ADM phase space, too. The seminal work in this direction can be found 
in \cite{Isham_1984b,Isham_1984a}. There, the reduced ADM phase space is extended by four embedding fields and their canonical momenta, while lapse and shift are still treated as Lagrange multipliers. Furthermore, a Gaussian condition for the metric is introduced in \cite{Isham_1984a}, so that the metric components $g_{00}$, $g_{0a}$ and their momenta are not elementary variables of the extended phase space. An extension of the ADM phase space that includes lapse and shift and their canonical momenta as well as the embedding fields together with their canonical momenta was introduced in \cite{Kouletsis_2008}. The difference to the extended phase space introduced in this article is that in \cite{Kouletsis_2008}, in addition to the embedding fields, all components of the spacetime metric and their momenta are part of the extended phase space, expressed by the spatial metric, lapse and shift and their momenta, while here, in addition to the embedding fields, the spatial metric, the normal vector field $n^\mu$ and the co-vector field $t_\mu$ encoding the time flow are included. In the latter case, lapse and shift and thus also the metric components $g_{00}$, $g_{0a}$ are only derived quantities, which are expressed as functions or functionals of the different choice of elementary phase space variables. As already mentioned, our construction is primarily motivated by the idea to translate covariant gauge fixing conditions into a suitable extension of the ADM phase space. An alternative approach for translating the harmonic gauge condition into the canonical theory can be found in the seminal work in \cite{Kuchar_1991}, where the reduced ADM phase space is extended by four scalar fields as well as four additional Lagrange multipliers that impose the harmonic gauge condition for the first four scalar fields. In this case, one imposes the harmonic gauge condition by means of a matter reference frame, while in the present work no additional matter couplings are considered. This is motivated by our purpose of linking edge modes to the geometrical clocks used in \cite{Fahn_2022}, which are constructed entirely from the gravitational sector without any contributions of additional scalar fields. Thus, the way the harmonic gauge condition is implemented here and in \cite{Kuchar_1991} differs by the number of physical degrees of freedom since  \cite{Kuchar_1991} adds matter fields, we do not.

\subsection{Boundary Currents and Charges}\label{sec:boundaryCurrents}

\noindent In Section ~\ref{sec:covariantPS}, we identified the boundary contributions to the pre-symplectic two-form. It is because of these boundary contributions that otherwise unphysical gauge directions become physical at the boundary. In the following, we compute the resulting charges for the gauge symmetries in linearized gravity.

Research on boundary charges and the symmetry algebra they generate is a large and active field, with many distinct extensions of the classical phase space, and various boundary conditions imposed (see e.g.~\cite{Ciambelli_2022b, Freidel_2023} for recent reviews). Here, we obtain the charges by integrating the  linearized currents over the two-dimensional boundary $\partial \Sigma$. This yields, first of all, the \emph{linearized momentum}
\begin{equation}
\delta P_\mu \equiv  \oint_{\partial\Sigma}\delta\boldsymbol{P}_\mu =\frac{1}{8\pi G}\oint_{\partial\Sigma}\big[(\ast-\gamma^{-1})\delta\Delta\big]_{\mu\nu}\wedge\di X^\nu_o.\label{eq:linMom1}
\end{equation}
Notice an important subtlety here. While the momentum current \eref{eq:Pcurr} carries a dependence on the Barbero--Immirzi parameter $\gamma$, the total charge is independent of $\gamma$. This follows from the linearized torsionless equation---the constraint $\delta T^\alpha=0$---which implies that $\di [\delta f^\mu]+\ou{\delta\Delta}{\mu}{\nu}\wedge\di X^\nu_o=0$. In fact, by applying Stokes's theorem, we quickly see the second term in the integral \eref{eq:linMom1} indeed vanishes,
\begin{equation}
\frac{1}{8\pi \gamma G}\oint_{\partial\Sigma}\delta\Delta_{\mu\nu}\wedge\di X^\nu_o=-\frac{1}{8\pi \gamma G}\oint_{\partial\Sigma}\di[\delta f^\mu]=0,
\end{equation}
leaving us with a linearized momentum that is independent of $\gamma$:
\begin{equation}
\delta P_\mu = \frac{1}{8\pi G}\oint_{\partial\Sigma}[\ast\delta\Delta]_{\mu\nu}\wedge \di X^\nu_o.
\end{equation}

A similar observation holds for the global angular momentum. This charge is the asymptotic integral of the boundary spin current  $\delta\boldsymbol{S}_{\mu\nu}$ plus an orbital contribution, which is the cross product $(\delta\boldsymbol{P}\wedge X_o)_{\mu\nu}$. While both terms each separately depend on $\gamma$, the integral of their sum over the boundary $\partial\Sigma$ does not.
All terms proportional to $\gamma$ combine to form a total derivative whose integral over $\partial\Sigma$  will vanish. In fact, if we look at the term in $\delta\boldsymbol{S}_{\mu\nu}$ that depends on $\gamma$, we obtain after performing a partial integration
\begin{align}\
\oint_{\partial\Sigma}\di X^{[\mu}_o\wedge \delta f^{\nu]}&=-\oint_{\partial\Sigma} X^{[\mu}_o\di \delta f^{\mu]}=\oint_{\partial\Sigma} X^{[\mu}_o\ou{\delta\Delta}{\nu]}{\rho}\wedge\di X^\rho_o.
\end{align}
The term on the right hand side of this equation cancels the $\gamma$-dependent contribution from the cross product $\oint_{\partial\Sigma}(\delta\boldsymbol{P}\wedge X_o)_{\mu\nu}$. We obtain
\begin{align}
\delta J^{\mu\nu}&=\oint_{\partial \Sigma}\delta\boldsymbol{J}^{\mu\nu}=-\frac{1}{4\pi  G}\oint_{\partial\Sigma}\Big(X_o^{[\mu}\ou{[\ast\delta\Delta]}{\nu]}{\rho}\wedge\di X^\rho_o-\frac{1}{2}\ou{\epsilon}{\mu\nu}{\mu'\nu'}\di X^{[\mu'}_o\wedge\delta f^{\nu']}\Big),
\end{align}
with no more dependence on $\gamma$ in the surface integral.\smallskip

\noindent \emph{Connection to $SU(2)$ Ashtekar-Barbero variables.} Finally, let us write the energy and momentum in terms of the linearized  Ashtekar-Barbero variables, which are the densitized triad 
\begin{equation}
\tilde{E}_i{}^a=\mathrm{d}^3x\,\det(e)\,\uo{e}{i}{a}\label{eq:Edens-def}
\end{equation}
and the $SU(2)$ Ashtekar-Barbero connection 
\begin{equation}
\ac^i{}_a = \Gamma^i{}_a[e] +\gamma K^i{}_a,\label{eq:AC-def}
\end{equation}
where $\Gamma^i{}_a[e]$ is the $SU(2)$ spin connection and the $\mathfrak{su}(2)$-valued one-form $K^i{}_a$ is obtained from the extrinsic curvature $K_{ab}$ through $K^i{}_a= e^{ib}K_{ba}$. In Subsection \ref{sec:LorntzChoice}, we introduced a specific choice of Lorentz frames to connect the perturbations of the Ashtekar-Barbero variables with the perturbations of the spin connection  $\delta\ou{\omega}{\mu}{\nu}:=\varphi^\ast_\Sigma\delta\ou{\Delta}{\mu}{\nu}$. Upon choosing this gauge, we obtain
\begin{align}
\delta\ou{K}{i}{a}=\ou{\delta\omega}{i}{0a},\quad
\delta\ou{\Gamma}{i}{a}=\tfrac{1}{2}\ou{\epsilon}{ki}{j}\ou{\delta\omega}{j}{ka}.
\end{align}
In the following, any terms proportional to $\mathrm{d}X_o^0$ vanish, because we are integrating over a surface of constant time $T=X^0_o$. Thus, starting from \eref{eq:linMom1}, we obtain the linearized energy
\begin{align}
    \delta P_0 &= \frac{1}{16\pi G}\oint_{\partial \Sigma}  \epsilon_{ijk} \delta \omega^{jk}  = -\frac{1}{8\pi G}\oint_{\partial \Sigma} \delta \Gamma_i\wedge \di X_o^i\label{eq:energyAV}
\end{align}
where $\delta \Gamma^i$ is a function of the linearized weighted triads, namely (cf. (2.17) in \cite{Fahn_2022})
\begin{equation}
\delta \Gamma^i{}_{a} = \frac{1}{2}\mathrm{d} x^i_b\utilde{\epsilon}^{bcd}\left[2\delta^f_{[d}\partial^k_{a]}\delta_{cg}-\delta^f_c\partial^k_d\delta_{ag}+\delta_{da}\delta^f_c\partial^k_g\right]\partial_f\delta\tilde{E}_k{}^g.
\end{equation}
Similarly, we find the linearized momentum in terms of the Ashtekar-Barbero connection,
\begin{align}
    \delta P_i &= \frac{1}{8\pi G}\oint_{\partial \Sigma}\epsilon_{ijk0} \delta \omega^{k0}\wedge \mathrm{d}X_o^j
    = \frac{1}{8\pi G}\oint_{\partial \Sigma}\epsilon_{ijk}{\delta K^k}\wedge \mathrm{d}X_o^j \nonumber\\
    & = \frac{1}{8\pi G\gamma}\oint_{\partial \Sigma} \epsilon_{ijk}\delta A^j \wedge \di X_o^k.\label{eq:momentumAV}
\end{align}
where in the last line we simply added a term proportional to $\epsilon_{ijk}\delta \Gamma^j \wedge \di X_o^k$. This term vanishes when integrating over $\partial \Sigma$ by the above arguments. In Appendix \ref{app: ADM_EM}, we further show how these observables are equivalent to the negative value of the linearized ADM energy and momentum \cite{Arnowitt_1962,Arnowitt_1959} respectively.

\subsection{Effects on the Quantum Theory}
\label{sec:BIparameter}

\noindent The  considerations above show that the Barbero--Immirzi parameter drops out from the global ADM charges. This raises the question whether there are any observables that can differentiate between different values of $\gamma$. One obvious answer lies in \emph{higher multipoles}, obtained by weighting the momentum and spin currents with some angle-dependent test function before integrating them over the boundary two-sphere. This leads us to integrals of the form $\oint_{\partial \Sigma} \rho^\mu(\theta,\phi)\delta \boldsymbol{P}_\mu$ for some angle-dependent $\rho^\mu(\theta,\phi)$. While the term proportional to $\gamma^{-1}$, which contributes to this integral, is a total derivative, the total integral is not and the contribution from the Barbero--Immirzi parameter survives. At the classical level, the appearance of $\gamma$ in the definition of the boundary charges has no observable consequences. It amounts to a mere relabeling of the charges. At the quantum level, this is no longer true, cf.\ \cite{Wieland_2024a}. To be more precise, we need to distinguish between the charges $Q_\alpha$, which are inferred algebraically as the generators of the symmetries, and the theory-independent observables $O_\alpha$ that we can measure in the real world. While there is no need to make this distinction for global charges, such as the total momentum or angular momentum, the difference matters once we consider the higher multipoles. For such observables, the canonical charges will be a $\gamma$-dependent function of the theory-independent observables, i.e.\ $Q_\alpha = f(\{O_\alpha\}, \gamma)$. At the classical level, this map does not affect the possible range of values the physical observables $O_\alpha$ can take. The observables $O_\alpha$ can be always written in terms of the metric and its derivatives alone. There is no dependence on $\gamma$ at this stage. At the quantum level, however, we have to take a detour: we start from the algebra of the abstract gauge charges, quantize them, find the possible eigenvalues $\{Q_\alpha\}$ of $\hat{Q}_\alpha$, and then invert the relation between $\{O_\alpha\}$ and $\{Q_\alpha\}$ to determine the corresponding observables $O_\alpha(\{Q_\alpha\},\gamma)$ such that a $\gamma$-dependence can appear at the level of the physical observables. \smallskip

At the quantum level, we are aware of similar effects of the addition of the Barbero--Immirzi parameter, which could have obseravble consequences. For example, it enters the LQG quantization of the proper area of a surface \cite{Ashtekar_1996,Rovelli_1994,Thiemann_2007} and it determines an upper bound for the matter density in the very early universe \cite{Ashtekar_2011}. More recently, it was found that there is a potential upper bound for the radiated gravitational wave power, which depends on $\gamma$ as well \cite{Wieland_2024b}. We believe that the way $\gamma$ enters the spectra of physical observables and the emergence of such bounds can very likely be traced back to the mechanism outlined above. \smallskip

Moreover, when going to the quantum theory starting from the pre-symplectic form (100), one has to quantize the
reference fields and their conjugate momenta, in addition to the standard gravitational radiation degrees of freedom
(cf. \cite[Sec.~4]{Kabel_2023}). While the construction of the resulting quantum states is, at this point, purely formal, we can nevertheless use it to gauge, in the metaphorical sense, where the Barbero--Immirzi parameter might affect the quantum theory. At the linearized level, where we ignore the self-interaction of gravitons, we are in the comfortable situation that we know the algebra of gravitational Dirac observables exactly. Their Poisson commutation relations can be inferred immediately by taking the pull-back of the pre-symplectic two-form to the constraint hypersurface. This amounts to setting $\delta H_\mu$ and $\delta G_{\mu\nu}$ in \eref{eq:diffOmega} and \eref{eq:Om-pertrb-lrntz} to zero. This in turn  implies that the bulk reference frames disappear from the symplectic two-form on the physical phase space. The only surviving contribution in the bulk  is given in \eref{eq:Omlinrzd-rad}, from which we can infer the commutation relations between the Dirac observables themselves. At the quantum level, these become the standard creation and annihilation operators $(a_\pm(\vec{k}), \bar{a}_\pm(\vec{k}))$ for the two polarization modes of gravitational radiation. However, this is not the end of the story. There are additional boundary terms that do not vanish upon imposing the constraints in the bulk. These boundary terms determine a boundary symplectic potential for the edge modes $\delta X^\mu|_{\partial\Sigma}\equiv\delta  Q^\mu(\vartheta,\varphi)$ and $\delta\barlambda^{\mu\nu}|_{\partial\Sigma}\equiv\delta\chi^{\mu\nu}(\vartheta,\varphi)$. At the canonical level, this boundary data $(\delta  Q^\mu,\delta\chi^{\mu\nu})$ can be linked to the boundary condition implicitly needed to specify the Green's function involved in the linearized gauge conditions \eref{eq:tcond}, \eref{eq:xcond} and \eref{eq:Xi-gauge-cond-lin2}.\smallskip

At the quantum level, the resulting kinematical state space thus takes the form of a tensor product $\mathcal{H} = \mathcal{H}_{\mathrm{bulk}}\otimes \mathcal{H}_{\mathrm{bndry}}$ between a bulk and a boundary Hilbert space. Since we have, at this stage, already solved the linearized constraints at the classical level, the quantization is simple. The bulk Hilbert space $\mathcal{H}_{\mathrm{bulk}}$ is the ordinary Fock space constructed from the mode expansion of the linearized gravitational perturbations.  For the boundary Hilbert space $\mathcal{H}_{\mathrm{bndry}}$, we choose a functional Schrödinger representation. Assuming a configuration space representation for the reference fields, we consider states $\Psi[\delta Q^\mu,{\delta\chi}^{\mu\nu}]\in \mathcal{H}$, where the dependence on the bulk gravitational degrees of freedom is left implicit to simplify notation. Not all the states in $\mathcal{H}$ are also physical. There is a residual constraint that links the bulk and boundary data. 
The relation $\delta P_\mu = \varphi^\ast_{\partial\Sigma}(\ast - \gamma^{-1})\delta \Delta_{\mu\nu}\wedge \di X_0^\nu \equiv \delta H_\mu$ connects the boundary intrinsic momentum, which is the conjugate variable to $\delta Q^\mu(\vartheta,\varphi)$, to the radiative modes in the bulk.\footnote{There is a similar condition for the  $SO(1,3)$ internal Lorentz symmetry, see \cite{Kabel_2023}.} This condition turns into a constraint on $\mathcal{H} = \mathcal{H}_{\mathrm{bulk}}\otimes \mathcal{H}_{\mathrm{bndry}}$. Physical states are annihilated by this constraint, which leads to the bulk-boundary relational Schrödinger-like equation 
\begin{equation}
 \widehat{(\delta P)}_\mu(\vec{\zeta}) \Psi_{phys}[\delta Q^\mu,\delta\chi^{\mu\nu}] = -\I\hbar\frac{\delta}{\delta Q^\mu(\vartheta,\varphi) } \Psi_{phys}[ \delta Q^\mu,\delta\chi^{\mu\nu}] =  \widehat{\delta H}_\mu(\vec{\zeta}) \Psi_{phys}[\delta Q^\mu,\delta\chi^{\mu\nu}].\label{eq:SchEq}
 \end{equation}

The construction is analogous to a standard Schrödinger picture, where the states depend parametrically on time. Here, however, time becomes multi-fingered: there is one relational clock $\delta Q^\mu(\vartheta,\varphi)$ at every point of the asymptotic two-sphere boundary. Given two kinematical states $\Psi, \Psi'\in\mathcal{H}$, the physical inner product is defined through a  projector via $\langle \Psi|\Psi'\rangle_{phys} = \langle \Psi |\boldsymbol{P}\Psi'\rangle$. The projector itself is a formal delta distribution implementing the constraint $\delta P_\mu - \delta H_\mu = 0$,
\begin{equation}
\boldsymbol{P} = \int_{[\R^4]^{S_2}}\mathcal{D}[N]\exp\left(\frac{\I}{\hbar}\oint_{S_2}N^\mu(\delta P_\mu - \delta H_\mu)\right).\label{eq:Prjectr}
\end{equation}
Consider now a quantum superposition of states, peaked around a particular configuration of the boundary reference fields. Since quantum theory is linear, we expect that these states can be written as a formal quantum superposition
\begin{equation}
\Phi=\sum_i\Phi_i[\delta Q_i,\delta \chi_i]\otimes|\delta Q_i,\delta \chi_i\rangle.
\end{equation}
Consider then a reference state 
\begin{equation}
\Psi=\Psi_o[\delta Q_o,\delta \chi_o]\otimes|\delta Q_o,\delta \chi_o\rangle.
\end{equation}
Formally, the inner product between physical states evaluates to
\begin{equation}
\langle\Psi,\Phi\rangle_{\mtext{phys}}=\sum_i\big\langle\Psi[\delta Q_o,\delta \chi_o]\big|\big(U^{\mathrm{BMS}}_{i\to 0}\Phi_i\big)[\delta Q_i,\delta \chi_i]\big\rangle_{\mtext{bulk}}.\label{eq:QRFtrafo}
\end{equation}
The inner product on the right hand side is simply the ordinary Fock inner product on $\mathcal{H}_{\mathrm{bulk}}$. The transformation between the two states is the formal matrix element of the projector 
\begin{align}
U^{\mathrm{BMS}}_{i\rightarrow 0} &= \langle \delta Q_o,\delta \chi_o| \boldsymbol{P} |\delta Q_i,\delta \chi_i\rangle.\label{eq:BMStrafo}
\end{align}
This  equation can be read in two ways. From the perspective of perturbative gravity,  $U^{\mathrm{BMS}}_{i\rightarrow 0}$ is a representation of the Bondi--Metzner--Sachs (BMS) group \cite{Sachs_1962,Bondi_1962,Ashtekar_1987,Ashtekar_1978} on the radiative phase space that maps the asymptotic frame $\{\delta Q_i,\delta\chi_i\}$ into $\{\delta Q_o,\delta\chi_o\}$. This makes it a primary element of the theory and the introduction of $\boldsymbol{P}$ is a mere rearrangement of equations. However, it could very well be  that in a full theory of quantum gravity, the logic is reversed and $U^{\mathrm{BMS}}_{i\rightarrow 0}$ is a derived quantity that only exists for sharply-peaked semi-classical states. In either case, we can notice that the sum $\sum_i \big(U^{\mathrm{BMS}}_{i\to 0}\Phi_i\big)[\delta Q_i,\delta\chi_i]$ implements a different such transformation depending on the state $i$ of the reference field and can thus be seen as a \emph{quantum-controlled} symmetry transformation, as commonly employed in the literature on quantum reference frames (see e.g.,~\cite{Giacomini_2017_covariance, Vanrietvelde_2018a, delaHamette_2021_perspectiveneutral, Castro_Ruiz_2021, Kabel_2024}).

We can now see that for \emph{constant translations} (i.e.\ translating the coordinate fields by the same amount $\delta Q_i^\mu-\delta Q_o^\mu = \text{const.}$ everywhere on the corner two-sphere), the effect of the Barbero--Immirzi parameter vanishes in the same way as it did for the global Poincaré charges since we are integrating uniformly over the linearized charge in \eref{eq:Prjectr}. However, we can also consider \emph{point-wise}, i.e.~angle-dependent, translations. In this case, as above, the integrand can no longer be written as a total derivative and the transformation remains non-trivial. So, while the global translations, associated to the global charges, remain unaffected by the introduction of $\gamma$, its effects likely survive when considering point-wise (quantum) reference frame transformations.

\section{Conclusions}
\label{sec:Conclusions}
\noindent 

\noindent In quantum gravity, all model building involves a choice of   reference frames. The strategy and the way in which these are chosen may differ from one approach to another. In this work, we began by exploring two different strategies for building such reference frames: the canonical ADM formalism and the covariant phase space formalism. This was motivated by recent results on geometrical clocks within the canonical framework \cite{Fahn_2022} and related research on edge modes for the diffeomorphism symmetry, which appeared in \cite{Kabel_2023}  within the covariant phase space formalism.\smallskip

To connect the two approaches, we introduced an extension of the reduced ADM phase space. The extension consists of additional coordinate fields $X^\mu$, the vector field $n^\mu$ normal to the Cauchy hypersurface, and the time-flow co-vector field  $t_\mu$, together with their conjugate momenta. This extension allowed us to impose covariant harmonic gauge fixing conditions on the extended 
ADM phase space. In this way, we could relate gauge-fixed coordinate fields on the covariant phase space to gravitational degrees of freedom in the extended ADM phase space. On the reduced ADM phase space, our construction gives rise to a gauge-fixing condition for the Hamiltonian and spatial  diffeomorphism constraints. Working in the regime of linearized gravity, we demonstrated that---upon choosing appropriate gauge fixing conditions for the normal vector field $n^\mu$ and the time-flow vector field $t^\mu$---the harmonic gauge fixing condition defines the same reference fields as the ones employed through the choice of particular geometrical clocks, constructed from Ashtekar--Barbero variables, see \cite{Fahn_2022}. Consequently, the linearized coordinate fields $\delta X^\mu$, used in the covariant phase space formalism, can be seen as a gauge-unfixed version of geometrical clocks identified in \cite{Fahn_2022}, constructed within the canonical framework. At the level of the canonical ADM phase space, the physical edge modes that appear in the the covariant phase space formalism reappear as boundary conditions required for defining the Green's function that appears in the definition of geometrical clocks through the covariant harmonic gauge fixing condition. This link warrants further investigation.\smallskip

After discussing the relation of our extended ADM phase space construction to related work, we further considered the implications of including the Barbero-Immirzi parameter $\gamma$ in the context of the covariant phase space formalism. Starting from the first-order Hilbert--Palatini action, we studied how the inclusion of $\gamma$ affects the symplectic structure and boundary modes. We derived the pre-symplectic two-form in the general and linearized regime and determined the $\gamma$-dependence of the boundary currents, conjugate to the edge modes for the diffeomorphism and internal Lorentz symmetry. While this dependence disappears in the expression for the global charges, it shows up at higher $Y_{\ell m}$-multipoles, which can affect physical observables at the quantum level. A similar effect can be found when considering boundary symmetry transformations: constant Poincaré translations of the asymptotic boundary two-sphere do not carry any dependence on $\gamma$, but the angle-dependent super-translations do.\smallskip

Our work provides  new insight on how to connect two different ways for modeling reference frames in general relativity: the canonical ADM approach and the covariant phase space formalism.  We achieved this match in the regime of linearized gravity and for a particular choice of geometrical clocks. We found that the geometrical clocks on the ADM phase space instantiate a special case of the harmonic gauge fixing at the covariant level. It would be interesting to explore other gauge fixing conditions in this spirit. As discussed in Section \ref{sec:ADMphasespace} as well as Section \ref{sec:coordConditions}, the harmonic gauge condition leads to a non-commutativity of the reference fields $X^\mu$. This agrees with the results for the geometrical clocks in \cite{Fahn_2022}. The appearance of this non-commutativity is expected, because the components of the harmonic gauge condition have non-vanishing Poisson brackets among each other.
As shown in \cite{Fahn_2022}, the geometrical clocks can be extended to mutually commuting (Abelian) clocks. This is achieved through a dual observable map that adds terms that are linear in the constraints to the non-commuting geometrical clocks and can be applied order by order in perturbation theory and thus extends beyond the linearized regime. An open question is whether this procedure can be carried over to covariant gauge fixing conditions. If it does, we would have a covariant modification of the harmonic gauge condition that would reproduce the \emph{mutually commutating} geometrical clocks of \cite{Fahn_2022}.

A related question is how to impose the covariant gauge fixing condition already at the level of the Lagrangian, by making appropriate modifications to the action, as done, for instance, in \cite{Kuchar_1991}, albeit using matter reference frames. It would be very interesting indeed to explore the resulting bulk and boundary field theory for various boundary conditions from the perspective of the canonical ADM formalism.\smallskip

Another intriguing direction for future research concerns the effects of the Barbero--Immirzi parameter $\gamma$ at the quantum level. We saw that it affects the boundary charges for super-translations, derived from the boundary currents, at both the classical and the quantum level. We argued that this deformation can affect physical observables. To derive  observable consequences, we need to determine the spectra of the boundary charges and understand them in terms of more standard observables that characterize the gravitational waves as measured by an asymptotic observer. Since the $\gamma$-term breaks parity, we may expect that the inclusion of $\gamma$ will affect the two polarizations differently, thus creating a fundamental asymmetry between left-handed and right-handed modes.\smallskip

At the quantum level, reference fields also have an exciting use in the definition of entanglement entropy for subregions in quantum field theory on curved spacetime. This is suggested by a number of recent works (e.g.~\cite{Witten_2021, Chandrasekaran_2022, Jensen_2023}) arguing that the addition of an {observer} degree of freedom reduces the von Neumann algebra of subregion observables from type III to type II, which allows for the definition of density matrices and entanglement entropy of spacetime subregions. The close connection of this observer degree of freedom to edge modes and quantum reference frames in various forms has been pointed out in several recent works \cite{Fewster_2024, deVuyst_2024, Ahmad_2024}. When included in the quantum theory, the linear perturbations of the coordinate fields $\delta X^\mu$  or an appropriate subset thereof could serve as a natural candidate for such an observer degree of freedom.\smallskip

Finally, it would be important to study how far our results carry over to the non-perturbative regime. Choosing geometrical clocks at the full non-linear level poses a far greater challenge if we are also interested to take the model to the quantum theory. This difficulty stems primarily from a substantially more complicated observable algebra. Furthermore, we would expect that the neat split of the kinematical variables into radiative modes and (gravitational) reference frames will, in general, not persist beyond the linear regime. Instead, we expect that these modes would likely be mixed in a non-perturbative treatment once the full non-linearity of general relativity is taken into account. This also carries over to the quantum theory when considering non-perturbative quantization methods, such as those applied in loop quantum gravity. A further intriguing avenue for future research involves examining this choice of (quantum) reference frames in the context of Fock representations of non-perturbative quantum gravity, as recently introduced in \cite{Thiemann_2024b,Thiemann_2024a}.\smallskip

\emph{Acknowledgments.}  WW  acknowledges that his contribution to this research was funded through a Heisenberg fellowship of Deutsche Forschungsgemeinschaft (DFG, German
Research Foundation)---543301681. VK acknowledges support through a DOC Fellowship of the Austrian Academy of Sciences. 
KG and WW thank Perimeter Institute for Theoretical Physics for hospitality and support. This research was supported in part by Perimeter Institute for Theoretical Physics. Research at Perimeter Institute is supported by the Government of Canada through the Department of Innovation, Science and Economic Development and by the Province of Ontario through the Ministry of Colleges and Universities. In addition,  WW sincerely acknowledges support for conducting this work from Edward Wilson-Ewing, Saeed Rastgoo, and Laurent Freidel during his recent visit to Canada. KG thanks IQOQI-Vienna for their kind hospitality. VK thanks FAU Erlangen-Nürnberg for their kind hospitality.

\bibliography{refQGRF}

\pagebreak
\appendix
\addtocontents{toc}{\string\tocdepth@munge}
\section{Comparison with existing work: detailed computations}\label{app:CompPons}
In this appendix we provide some more details on the computations that are presented in subsection \ref{subsec:relADMphaseSpace}, where mainly the results are discussed.

First, we compute the action of the improved diffeomorphism generator $\boldsymbol{G}_\xi$ defined in \eqref{eq:Gdef} on the lapse function. We obtain
\begin{align}\nonumber
\big\{N,\boldsymbol{G}_\xi\big\}&=-\xi^\mu\nabla_\mu (n^\nu t_\nu)-n_\mu[t,\xi]^\mu+(\mathcal{L}_\xi n_\mu)\ou{h}{\mu}{\nu}t^\nu=\\
\nonumber &= (\mathcal{L}_\xi n_\mu)n^\mu n_\nu t^\nu=-Nn^\mu\mathcal{L}_\xi n_\mu=\\
&=-Nn^\mu\nabla_\mu\xi^\nu n_\nu=Nn^\mu\nabla_\mu\xi_\perp+N\xi_\parall^a\beta_a.\label{eq:GxiN1A}
\end{align}
Taking into account that $t^\mu=N n^\mu+\ou{h}{\mu}{a}N^a$ and the definition of the acceleration vector $\beta^a$, see \eref{eq:acclrtn},  \eref{eq:GxiN1A} translates into
\begin{equation}
\big\{N,\boldsymbol{G}_\xi\big\}=\dot{\xi}_\perp-N^a D_a\xi_\perp+\xi^a_\parall D_a N,\label{eq:GxiN2A}
\end{equation}
where $\dot{\xi}_\perp=-t^\nu\partial_\nu(n^\mu \xi_\mu)=\{\xi_\perp,\boldsymbol{H}\}$. 

Second, we consider how the diffeomorphism generator $\boldsymbol{G}_\xi$ acts on the shift vector. Going back to the definition of the improved generator \eref{eq:Gdef}, we get
\begin{align}
\nonumber \ou{\di X}{\mu}{a}\big\{\boldsymbol{G}_\xi,N^a\big\}&=\ou{\di X}{\mu}{a}\{ h^{ab},\boldsymbol{G}_\xi\}\partial_bX^\nu t_\nu+\di X^{\mu a}\partial_a\xi^\nu t_\nu+\di X^{\mu a}\partial_aX^\nu\xi^\rho\partial_\rho t_\nu+\ou{h}{\mu}{\nu}[t,\xi]^\nu=\\
\nonumber&=\mathcal{L}_\xi(\ou{\di X}{\mu}{a} N^a)+n^\mu n_\rho(\mathcal{L}_\xi h^{\rho\nu})t_\nu+(\mathcal{L}_\xi h^{\mu\rho})n_\rho n^\mu t_\mu+\ou{h}{\mu}{\nu}[t,\xi]^\nu=\\
\nonumber&=\ou{h}{\mu}{\nu}\mathcal{L}_\xi(\ou{\di X}{\nu}{a} N^a)-\ou{h}{\mu}{\nu}\mathcal{L}_\xi t^\nu+Nh^{\mu\rho}\mathcal{L}_\xi n_\rho=\\
\nonumber&=-\ou{h}{\mu}{\nu}\mathcal{L}_\xi(Nn^\nu)+Nh^{\mu\rho}(\xi^\sigma\nabla_\sigma n_\rho+\nabla_\rho\xi^\sigma n_\rho)=\\
\nonumber&=\ou{h}{\mu}{\nu}[Nn,\xi_\perp n+\xi_\parall]^\nu+N \di X^{\mu a}(\xi_\perp\beta_a-D_a\xi_\perp).
\end{align}
In here, $\mu,\nu,\dots$ are abstract space time indices, whereas $a,b,c,\dots$ are abstract indices for tensor fields intrinsic to $\Sigma$. Now, since $Nn^\mu=t^\mu-\ou{\di X}{\mu}{a}N^a$, we obtain
\begin{equation}
\big\{N^a,\boldsymbol{G}_\xi\big\}=\dot{\xi}^a_\parall-N^bD_b\xi^a_\parall+\xi^b_\parall D_bN^a+h^{ab}(\xi_\perp D_bN-ND_b\xi_\perp),
\end{equation}
where 
\begin{equation}
\dot{\xi}^a_\parall=\big\{h^{ab}\partial_bX^\mu\xi_\mu(X),\boldsymbol{H}\big\}=\uo{\di X}{\mu}{a}\mathcal{L}_t(\ou{\di X}{\mu}{b}\xi^b_\parall).
\end{equation}
\section{Boundary Currents in the Full Theory}\label{app: BoundaryCurrents}

In this appendix, we derive the boundary momentum and spin current for an asymptotic boundary. In particular, we consider spatial infinity $i_o$, obtained in the limit $r=\sqrt{X_iX^i} \to \infty$, with asymptotic falloff conditions\footnote{Note that the tensor components $\ou{\Delta}{\mu}{\nu\rho}$ and $\ou{f}{\mu}{\nu}$ have a faster falloff. The standard conditions are $\ou{\Delta}{\mu}{\nu\rho}=\mathcal{O}_-(r^{-2})$ and $\ou{f}{\mu}{\nu}=\mathcal{O}_+(r^{-1})$, where $\mathcal{O}_\pm$ indicates even (odd) parity, e.g.\ $\ou{f}{\mu}{\nu}(T,\vec{X})=\ou{f}{\mu}{\nu}(T,-\vec{X})+\mathcal{O}(r^{-2})$.}, $\Delta^{\mu\nu}=\mathcal{O}(r^{-1})$ and $f^\mu=\mathcal{O}(r^0)$, as well as $\bbvar{d}X^\mu=\mathcal{O}(r^0)$. Starting from the last line of the boundary symplectic two-form \eref{eq:Omphysbound}, we obtain

\begin{align}\nonumber
\frac{1}{16\pi G}&\oint_{\partial\Sigma\rightarrow i_o}\bigg[2\bbvar{X}_\alpha e_\beta\wedge(\ast-\gamma^{-1})\left(\bbvar{D}A^{\alpha\beta}+\bbvar{X}\hook F^{\alpha\beta}\right)-\bbvar{X}_\alpha\bbvar{X}_\beta(\ast-\gamma^{-1})F^{\alpha\beta}\bigg]=\\\nonumber
&=\frac{1}{16\pi G}\oint_{\partial\Sigma\rightarrow i_o}\bigg[2\bbvar{d}X_\mu \di X_\nu\wedge(\ast-\gamma^{-1})\bbvar{d}\Delta^{\mu\nu}-\bbvar{d}X_\mu\bbvar{d}X_\nu(\ast-\gamma^{-1})\di\Delta^{\mu\nu}\bigg]=\\
\nonumber
&=\frac{1}{8\pi G}\oint_{\partial\Sigma\rightarrow i_o}\bbvar{d}X_\mu \bigg[\di X_\nu\wedge(\ast-\gamma^{-1})\bbvar{d}\Delta^{\mu\nu}+\bbvar{d}(\di X_\nu)(\ast-\gamma^{-1})\Delta^{\mu\nu}\bigg]=\\
&=-\oint_{i_o}\bbvar{d}X^\mu\bbvar{d}\boldsymbol{P}_\mu,\label{eq:asymptoticOmegabound}
\end{align}

where we introduced the boundary momentum current

\begin{equation}
\boldsymbol{P}_\mu=\frac{1}{8\pi G}(\ast\Delta-\gamma^{-1}\Delta)_{\mu\nu}\wedge\di X^\nu.
\end{equation}

Note that, while the coordinates $X^\mu$ are of order $\mathcal{O}(r)$, their variation $\bbvar{d}X^\mu$ needs to be of order $\mathcal{O}(r^0)$ in order to obtain a finite expression in \eqref{eq:asymptoticOmegabound}. This is true for angle-dependent translations. In this case, there is a translation that shifts the coordinates at each point on the boundary two-sphere by a vector, which is $\mathcal{O}(r^0)$, such that the resulting variation $\bbvar{d}X^\mu$ is independent of $r$. In the same way, we obtain the intrinsic gravitational spin, which is conjugate to the $SO(1,3)$ spin frames at the asymptotic boundary.
It contains a contribution from the vacuum, which diverges quadratically in $r$. Subtracting this contribution, we obtain
\begin{equation}
\boldsymbol{S}_{\mu\nu}=\frac{1}{4\pi G}\left(\tfrac{1}{2}\ou{\epsilon}{\rho\sigma}{\mu\nu}-\gamma^{-1}\delta^{[\rho}_{\mu}\delta^{\sigma]}_{\nu}\right)\,\di X_{[\rho}\wedge f_{\sigma]}.
\end{equation}
The momentum and spin current together form the total angular momentum current 
\begin{equation}
\boldsymbol{J}_{\mu\nu}=2 \boldsymbol{P}_{[\mu}X_{\nu]}+\boldsymbol{S}_{\mu\nu}.\label{eq:Jchrg1}
\end{equation}
Taking into account the linearized Einstein equations $\di\ou{(\ast\Delta)}{\mu}{\nu}\wedge\di X^\nu=0$ and $\di f^\mu+\ou{\Delta}{\mu}{\nu}\wedge \di X^\nu=0$, it is immediate to check that the currents are conserved. For the momentum current $\boldsymbol{P}_\mu$, this follows directly from $\di\ou{(\ast\Delta)}{\mu}{\nu}\wedge\di X^\nu=0$ and the linearized Bianchi identity $\di\ou{\Delta}{\mu}{\nu}\wedge\di X^\nu=0$.

For the total angular momentum current $\boldsymbol{J}_{\mu\nu}$, the calculation is easier when using self-dual variables. With respect to two-component Weyl spinor inices $A,B,C,\dots=0,1$, the selfdual part of the angular momentum charge \eref{eq:Jchrg1} is
\begin{align}\nonumber
\boldsymbol{J}_{AB}&=\frac{\I}{8\pi\gamma G}(\gamma+\I)\Delta_{C(A}\wedge\di X^{C\bar{C}}X_{B)\bar{C}}+
\frac{\I}{8\pi\gamma G}(\gamma-\I)\ou{\bar\Delta}{\bar{C}}{\bar{D}}\wedge\uo{\di X}{(A}{\bar{D}}X_{B)\bar{C}}+\\
&\qquad+\frac{\I}{8\pi\gamma G}(\gamma+\I)\uo{\di X}{(A}{\bar{C}}\wedge f_{B)\bar{C}}.
\end{align}
In these variables, the linearized Einstein equations are
\begin{align}
&\di\Delta_{AB}\wedge \ou{\di X}{B}{\bar{A}}=0,\\\
&\di f^{A\bar{A}}+\ou{\Delta}{A}{C}\wedge\di X^{C\bar{A}}+\ou{\bar\Delta}{\bar A}{\bar{C}}\wedge\di X^{A\bar{C}},
\end{align}
where $\Delta_{AB}=\Delta_{BA}$ is the perturbation of the self-dual connection. It is then straightforward to show that
\begin{equation}
\di \boldsymbol{J}_{AB}=0.
\end{equation}

\section{Relation of Boundary Currents to ADM Energy and Momentum}\label{app: ADM_EM}

In this appendix, we show that the linearized boundary currents Eqs.~\eref{eq:energyAV} and \eref{eq:momentumAV} are equivalent to the negative value of the linearized ADM energy and momentum \cite{Arnowitt_1962,Arnowitt_1959} respectively. To see the equivalence for the energies, we first write out \eref{eq:energyAV} in terms of $\delta E^i_{\;\; a}$,
\begin{align*}
    \delta P_0 &= -\frac{1}{8\pi G}\oint_{\partial \Sigma} \delta \Gamma_i\wedge \di X_o^i = -\frac{1}{8\pi G}\oint_{\partial \Sigma} dS^c \epsilon_c^{\;\; ab} \delta \Gamma^i_{\;\; a}\partial_{b i}\\
    &= \frac{1}{16\pi G} \oint_{\partial \Sigma} dS^c \epsilon_c^{\;\; ab} \left(  -2\epsilon_{b\tilde{c}}^{\;\;\;\; \tilde{b}} \partial^l_{[a} \partial_{\tilde{b}]} (\delta E^{\tilde{c}}_{\;\; l}) - \epsilon_b^{\;\; d\tilde{b}} \partial_d^l \delta_{a\tilde{c}}\partial_{\tilde{b}} (\delta E^{\tilde{c}}_{\;\; j}) + \epsilon_{ba}^{\;\;\;\; \tilde{b}} \partial_{\tilde{b}}(\delta E^{\tilde{c}}_{\;\; l})  \right)\\
    &= -\frac{1}{8\pi G}\oint_{\partial \Sigma} dS^c \partial^l_c \partial_{\tilde{c}}(\delta E^{\tilde{c}}_{\;\; l}) =  -\frac{1}{8\pi G}\oint_{\partial \Sigma} dS^a \partial^l_c (E^i_{\;\;a})^{(0)} \partial_b(\delta E^{\tilde{c}}_{\;\; l}).
\end{align*}
This can be rewritten in terms of the non-weighted triads using $\delta E_i^a = \utilde{\epsilon}^{abc} \epsilon_{ijk} \partial^j_b \delta e^k_c$. Using this, we find that
\begin{align*}
    \delta P^0 &= +\frac{1}{8\pi G}\oint_{\partial \Sigma} dS^a \partial^l_c \partial^i_a \partial_b(\epsilon^{bcd}\epsilon_{ijk} \partial^j_c\delta e^k_d) \\
    &= -\frac{1}{8\pi G}\oint_{\partial \Sigma} dS^a \left( \partial^b(\partial_{bi} \delta e^i_{a}) - \partial_a (\partial^b_i \delta e^i_b) \right)\\
    &= -\frac{1}{16\pi G}\oint_{\partial \Sigma} dS^a(\partial^b \delta h_{ab} - \partial_a \delta h^b_{\;\; b}) = -\delta E_{ADM}.
\end{align*}

Note that, in going to the last line, we used that $\delta h_{ab} = 2 \eta_{ij}\partial^i_{(a}\delta e^j_{b)}$ and that the term involving the anti-symmetric part $\delta_{ij}\partial^i_{[a}\delta e^j_{b]} = 0$ vanishes. The latter can be seen by encoding the anti-symmetric part in a one-form $\delta \omega$ such that $\delta \omega$ such that $\epsilon_{acb}\delta\omega^c:=\delta_{ij}\partial^i_{[a}\delta\ou{e}{j}{b]}$. Then, 
\begin{equation}
\oint_{\partial\Sigma}d^2S^a\partial^b(\delta_{ij}\partial^i_{[a}\delta\ou{e}{j}{b]})=\oint_{\partial\Sigma}d S^a\epsilon_{abc}\partial^b\omega^c=
\oint_{\partial\Sigma}\di\omega=0,
\end{equation}

Next, let us turn to the momentum given in \eref{eq:momentumAV}, which can be rewritten as
\begin{align*}
    \delta P_i &= \frac{1}{8\pi G}\oint_{\partial \Sigma} dS^c \epsilon_c^{\;\; ab}\epsilon_{ijk} \delta K^j_{\;\; a}\partial_{b}^k\\
    &= \frac{1}{8\pi G}\oint_{\partial \Sigma} dS^c (\partial_{ci}\partial^a_j - \partial_{cj}\partial^a_i)\delta K^j_{\;\; a}\\
    &= \frac{1}{8\pi G}\oint_{\partial \Sigma} dS^c (\partial_{ci}\delta K^a_{\;\;a} - \partial^a_i \delta K_{ca})\\ &= -\frac{1}{8\pi G}\oint_{\partial \Sigma} dS^b(\delta K_{ab} - (\delta K) \delta_{ab})\partial^a_i = -\partial^a_i(\delta P_{\mathrm{ADM}})_a.
\end{align*}
Note that the fact that the conserved charges are not equivalent to the linearized ADM energy and momentum themselves but rather to their negative value is consistent with our convention for the constraint $\delta\boldsymbol{P}_\mu + \delta \boldsymbol{H}_\mu =0$ (cf. \cite{Struckmeier_2005}). 

\end{document}